\documentclass[structabstract]{aa}
\usepackage{natbib}
\bibpunct{(}{)}{;}{a}{}{,}
\usepackage{txfonts}
\usepackage{graphicx}
\usepackage{longtable}
\newcommand{\Teff} {$T_\mathrm{eff}$}
\title{The masses, and the mass discrepancy of O-type stars}
\author{Carsten~Weidner\inst{\ref{inst1}}\and Jorick S.~Vink\inst{\ref{inst2}}} 
\institute{Scottish Universities Physics Alliance (SUPA), School of Physics and
  Astronomy, University of St Andrews, North Haugh, St Andrews, Fife
  KY16 9SS, UK \email{cw60@st-andrews.ac.uk}\label{inst1}
\and
Armagh Observatory, College Hill, Armagh BT61 9DG, UK
\email{jsv@arm.ac.uk}\label{inst2}
}

\begin{document}
\date{Received 2010 / Accepted 2010}

\label{firstpage}

\abstract{The ``mass discrepancy'' in massive O stars represents a 
long-standing problem in stellar astrophysics with far-reaching 
implications for the chemical and dynamical feedback in galaxies.}
{Our goal is to investigate this mass discrepancy by comparing state-of-the-art
model masses with model-independent masses determined from eclipsing binaries.}
{Using stellar evolution models and a recent calibration of 
stellar parameters for O-star spectral sub-classes, we 
present a convenient way to convert observed solar metallicity O
star spectral types into model masses, which we subsequently compare
to our dynamical mass compilation. We also derive similar
conversions for Large and Small Magellanic Cloud metallcities.} 
{We obtain a good agreement between model and dynamical masses, suggesting 
the long-standing problem of a systematic mass discrepancy problem
may have been solved. We also provide error ranges for the
model masses, as well as minimal and maximal age estimates for when
the model stars are in a given spectral type box.}
{}


\titlerunning{The masses of O stars}
\authorrunning{Weidner \& Vink}

\maketitle

\section{Introduction}
\label{sec:intro}

The most basic parameter of a star is its mass. Knowledge of this most
fundamental parameter is of utmost importance for basically all  
of astrophysics. For massive O stars, reliable
mass determinations have turned out to be particularly challenging.
For over two decades there has been a ``mass discrepancy'' where 
O-star masses derived from evolutionary models ($M_{\rm evol}$) were 
found to be systematically higher than those derived from stellar 
atmosphere analyses ($M_{\rm spec}$) by up to a factor $\sim$2
\citep{GL89,HKV92}.
Even two recent studies still report a
significant mass discrepancy for non-enriched O-type stars in the
Large Magellanic Cloud \citep[LMC,][]{MKE07} and the Milky Way
\citep[MW,][]{HNS10}.

Over the last few decades, four alternative methods to determine O star masses
have been developed:

\begin{itemize}
\item{} evolutionary masses ($M_{\rm evol}$)
\item{} spectroscopic masses ($M_{\rm spec}$)
\item{} wind masses ($M_{\rm wind}$)
\item{} dynamical masses ($M_{\rm dyn}$).
\end{itemize} 

In the first method, one places the luminosity (or absolute magnitude) 
and effective temperature (or colour) in a Hertzsprung-Russell diagram (HRD) 
and compares the positions of the stars with theoretical stellar evolution
models. 

The second way comprises the use of stellar 
spectroscopy: via the Stark broadening in spectral lines, one can derive 
log $g$ and subsequently the mass $M_{\rm spec}$. 
For O stars, this method is highly complex, because stellar winds 
have a severe influence on the underlying model atmospheric structure 
\citep[e.g.][]{GGK89,H91}. 

In the meantime, a third method to determine O star wind masses
$M_{\rm wind}$ had been put forward \citep{GL89,KHP92}. This method 
employs the radiation-driven theory \citep[e.g.][]{CAK75}, which relates 
the terminal wind velocity to the stellar escape
velocity. \citet{KHP92} and \citet{HKV92} 
found good agreement between their spectroscopic and wind masses, 
and suggested that the evolutionary masses were systematically too large. 
Although there was indeed no particular reason to expect that  
evolutionary masses should be correct -- given the large number of 
uncertainties in the underlying physical input
(e.g. mass loss, overshooting, rotation, and magnetic effects) -- the
evolutionary calculations seemed to reproduce the observed O-star
properties rather well \citep{HHB96}. Subsequent work by
\citet{BMM97}, who tried to derive masses from binary dynamics
($M_{\rm dyn}$), suggested that the evolutionary masses were at least
of the right order of magnitude, thereby challenging the spectroscopic
masses, which were significantly lower at the time.

The best argument to trust the spectroscopic masses was their independent 
agreement with wind masses based on radiation-driven wind theory. 
This was not always the case as in the 1990s, there was 
also a systematic discrepancy between mass-loss rates predicted by 
wind theory and observations \citep{LL93,PKH96}. This
situation changed when \citet{VDL00} presented new wind models
including multiple-scattering. These models no longer show the
systematic discrepancy with empirical rates\footnote{Although it is
  currently debated whether the absolute values of these mass-loss
  rates are of the right order of magnitude. Some recent studies have
  called for a fundamental reduction in O-star mass-loss rates as a
  result of wind clumping \citep{BLH05,FMP06}.}. 
Although the good agreement reached between these new mass-loss predictions 
and the empirical rates -- using the evolutionary rather than the spectroscopic 
masses -- could have been coincidental, the additional compatibility between 
evolutionary and dynamical masses, resulting in an agreement between
{\it three} methods, led to the suspicion that it was most likely the
spectroscopic masses that were the main culprit for the mass discrepancy.

\citet{LDH96} had already suggested that the neglect of line-blanketing could
cause $M_{\rm spec}$ to be underestimated, and subsequent improvements
resulted in a new calibration of Galactic O-star parameters by
\citet[][hereafter MSH05]{MSH05}. These state-of-the-art non-local
thermodynamic equilibrium (NLTE) models include both mass loss and
line-blanketing. 
In the meantime, the effects of stellar rotation were included in the 
Geneva evolutionary models \citep{MM03}, and below we will indeed
confirm that the evolutionary masses now agree with the spectroscopic
ones. This should be considered a major triumph for the formidable
task of including full Fe line-blanketing in the atmospheric models
\citep[e.g.][]{HLH03}.

Nevertheless, given that both the Geneva evolutionary masses and the
MSH05 calibration include the same \citet{VDL00} mass-loss rates,
even an agreement with the wind masses could be a coincidence
involving complex model interdependencies, and it is by no means certain
they should be correct. It thus remains crucial to check our model
masses against model-independent ones.

The only known model-independent masses so far are the dynamically
derived ones: $M_{\rm dyn}$. This method is only  
applicable to binary stars. Usually, careful determination of the orbital 
parameters of a system allows one to obtain the mass ratio of the two stars, 
as the orbit inclination relative to Earth is generally unknown. 
Only for eclipsing binaries, the inclination is well-enough 
constrained to be able to measure absolute stellar 
masses. Unfortunately, stars eclipsing each other are very rare, and the 
search for these systems comprises an important endeavor to calibrate
and verify evolutionary models. In this paper we provide a compilation
of dynamical masses derived from eclipsing binaries.
Only very few eclipsing binaries are known that composed of at least
one massive star\footnote{For the context of this 
publication massive stars are solely O stars that are thought to
have masses of about 16 $M_\odot$ and larger.}. All of these are
challenging to study because they are distant and, as most or probably all
massive stars are born in star clusters \citep{AM01,LL03,AMG07}, in
very crowded regions of the sky. 
Nonetheless, a growing sample of eclipsing O stars is known, and 
we intend to use these to calibrate a relation between the spectral type 
of an O star and its present-day mass. 
Such a relation allows us also to calibrate the masses of O stars that are not 
part of an eclipsing binary system, i.e. the vast majority!

In Sect.~\ref{sec:def} the spectral classification of O stars is discussed,
while in Sect.~\ref{sec:data} the mass determinations from eclipsing
binaries and spectroscopic masses are presented. The results of this
study follows in Sect.~\ref{sec:disc}, and a summary is provided in
Sect.~\ref{sec:summ}. The stellar evolution models used and our 
interpolation scheme for additional intermediate model masses is described in
appendix~\ref{app:evol}. Furthermore, a large table is provided in the
appendix~\ref{app:stevol}, which shows the spectral type evolution of
the different stellar models used (on-line only).  

\section{Spectral classification of massive stars}
\label{sec:def}
Traditionally, O star spectral types are divided into sub-types from O3 to
O9, with 0.5 steps (but without the sub-types O3.5 and O4.5). These 
are observationally defined by the relative strength of the HeI to HeII
lines \citep{CA71}. \citet{WHL02} additionally defined the early O2,
O2.5 and O3.5 subtypes, but because the O2 sub-type classification
involves the nitrogen (N) sequence rather than the He sequence, these
additional sub-types are not (yet) universally accepted. For instance,
MSH05 do not use them. For numerical simplicity, we employ the range
of sub-types O2 to O9.5, divided into bins of 0.5 width, and with
luminosity classes: V (main-sequence), III (giant), and I (supergiant).

\begin{figure}
\begin{center}
\includegraphics[width=8cm]{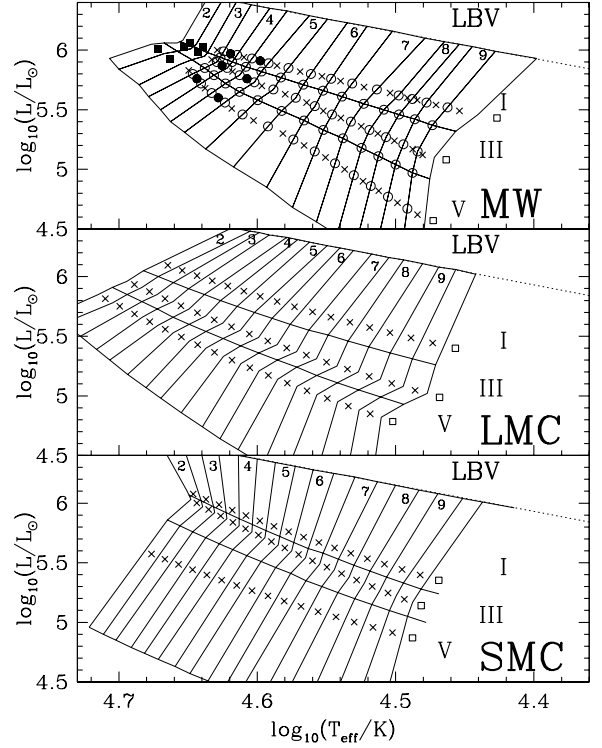}
%\vspace*{-2.0cm}
\caption{Definitions of the solar metallicity O star spectral types in
  the luminosity-temperature diagram derived from MSH05 ({\it panel
    MW}), the LMC metallicity ({\it panel LMC}) and the SMC
  metallicity ({\it panel SMC}) derived in this work. {\it Panel MW}
  shows the original MSH05 data as {\it crosses} 
  and the interpolated 3.5 and 4.5 subclasses as {\it filled
    circles}. Plotted as {\it filled boxes} are the extrapolated
  subtypes 2.5 and 2. The {\it open circles} connected by {\it solid
    lines} representes the interpolated grid that defines the $L$ and
  $T_\mathrm{eff}$ values for each spectral subtype. The luminosity
  classes are indicated as Roman numerals (I - supergiants, III -
  giants and V - dwarfs), while the spectral subtypes are shown by
  Arabic numerals. The upper limit ({\it dotted line}) is given by
  the hot edge for luminous blue variables (LBV) given in
  \citet{SVD04}. All interpolations are derived by calculating
  mean values in linear space. {\it Panel LMC} shows the  LMC
  metallicity grid derived from \citet{MKE07} and {\it panel SMC} the
  one for SMC metallicity obtained from \citet{HLH06}.}
\label{fig:sptypes}
\end{center}
\end{figure}

These sub-type--luminosity-classes (from now on spectral type
boxes) are defined by six vertices each, the four corner points and
the mean values between the central points of each box. Each vertice is
described by a luminosity and a \Teff. {\it Panel MW} of
Fig.~\ref{fig:sptypes} shows these boxes for the solar metallicity
grid. The central points of the boxes (marked with {\it crosses}) are
the new O star spectral type 
calibrations by MSH05. Table~\ref{tab:LTgrid} shows the values of the
vertices for the solar metallicity spectral boxes. As the subtypes 
O3.5 and O4.5 are not provided by MSH05, the corresponding values
(shown as {\it black dots} in {\it Panel A} of Fig.~\ref{fig:sptypes})
are derived by interpolation between O3 and O4, and O4 and O5,
respectively. For the O2 and O2.5 subtypes, which are not used by
MSH05, their results are extrapolated to these classes ({\it open
  circles} in {\it Panel A} of Fig.~\ref{fig:sptypes}). The following
fits to their theoretical \Teff~calibration are used to define the
\Teff~values for O2 and O2.5 for the three luminosity classes:
\begin{eqnarray}
\mathrm{Class~I}:& T_\mathrm{eff} =& 48598 - (ST * 2016) \nonumber\\
\mathrm{Class~III}:& T_\mathrm{eff}=& 49045 - (ST * 2034)\\
\mathrm{Class~V}:& T_\mathrm{eff}=& 50941 - (ST * 1978), \nonumber
\end{eqnarray}
were $ST$ is the spectral subtype. With these \Teff~values and MSH05 
Eqs.~(3), (4) and (5), the corresponding luminosities are calculated. The
upper limit for the supergiants is set by the empirical hot edge of
the Luminous Blue Variables (LBV) from \citet{SVD04}
$\log_{10}(L_\mathrm{LBV min}) = 2.2056 \cdot
\log_{10}(T_\mathrm{eff}) - 3.7737$. If a star is above 
this line it is regarded to be an LBV. The lower limit is somewhat
arbitrarily set as a parallel line to the luminosity class V, shifted
towards higher temperatures. The border between the O9.5 and B0
subclasses is set by the B0 definitions from \citet{ZCA09} for the
$T_\mathrm{eff}$ and the luminosities from \citet{SPM08}. Whenever a
star is earlier than O2, it is designated O2.0 If$^\ast$ in the
Tables~\ref{tab:Orot} to \ref{tab:Oz04} and \ref{tab:MM03z20}.

Naturally, such a scheme is not to be expected to be in full
compliance with how the spectral indices for O stars change with
\Teff. To compensate for this, a general error of 1000K is assumed 
for all \Teff~values used.

About one third of the stars with dynamical mass estimates
(Table~\ref{tab:dyn}) are located in the LMC. This dwarf galaxy has
a considerably lower metallicity than solar
\citep[$z_\mathrm{LMC}$~$\approx$~0.008,][]{VDB00} and therefore solar
metallicity spectral definitions and evolutionary models might not
represent these stars well. Especially the \Teff~scale of LMC
metallicity stars is well above solar metallicity ones
\citep{EC09}. As a comprehensive study of O type spectral classes like
MSH05 does not exist for LMC metallicity stars, the sample of LMC O
and early B-type stars of \citet{MKE07} is used to fit the following
relations between \Teff~and $\log_{10}(L)$ for LMC metallicity O stars.

\begin{eqnarray}
\mathrm{Class~I~(LMC)}:& T_\mathrm{eff} =& 50597 - (ST * 2197) \nonumber\\
\mathrm{Class~III~(LMC)}:& T_\mathrm{eff}=& 53713 - (ST * 2432)\\
\mathrm{Class~V~(LMC)}:& T_\mathrm{eff}=& 56143 - (ST * 2437). \nonumber
\end{eqnarray}
\begin{eqnarray}
\mathrm{Class~I~(LMC)}:& \log_{10}(L) =& 6.269 - (ST * 0.08698) \nonumber\\
\mathrm{Class~III~(LMC)}:& \log_{10}(L)=& 6.170 - (ST * 0.11850)\\
\mathrm{Class~V~(LMC)}:& \log_{10}(L)=& 6.073 - (ST * 0.12877). \nonumber
\end{eqnarray}
The resulting spectral type grid is shown in Table~\ref{tab:LTgrid}
and as {\it Panel LMC} of Fig.~\ref{fig:sptypes}.

While there are no eclipsing binaries in Table~\ref{tab:dyn} with SMC
metallicities \citep[$z_\mathrm{SMC}$~$\approx$~0.004,][]{VDB00}, a
recent \Teff~calibration for O stars in the SMC
does exist \citep{HLH06}. Their data are used to derive the following
\Teff~and luminosity relations in dependence of the spectral subtype.

\begin{eqnarray}
\mathrm{Class~I~(SMC)}:& T_\mathrm{eff} =& 48000 - (ST * 1857) \nonumber\\
\mathrm{Class~III~(SMC)}:& T_\mathrm{eff}=& 47456 - (ST * 1715)\\
\mathrm{Class~V~(SMC)}:& T_\mathrm{eff}=& 51660 - (ST * 2092), \nonumber
\end{eqnarray}

\begin{eqnarray}
\mathrm{Class~I~(SMC)}:& \log_{10}(L) =& 6.258 - (ST * 0.09048) \nonumber\\
\mathrm{Class~III~(SMC)}:& \log_{10}(L)=& 6.216 - (ST * 0.10763)\\
\mathrm{Class~V~(SMC)}:& \log_{10}(L)=& 5.750 - (ST * 0.08798). \nonumber
\end{eqnarray}

Like in the other two cases the spectral type grid is included in
Table~\ref{tab:LTgrid} and it is shown in {\it Panel SMC} of
Fig.~\ref{fig:sptypes}. As is visible in Fig.~\ref{fig:sptypes} and
Table~\ref{tab:LTgrid}, the SMC \Teff~grid is shifted to {\it lower}
temperatures compared to the LMC grid. The reason for this is non-trivial, 
and deserves a thorough analysis using SMC metallicity NLTE atmospheres, 
which is beyond the scope of this paper. 
Note that the \citet{HLH06} sample only includes one star with
a spectral type earlier than O4. This star is not included in the
relations, but the relations are used from O2 to O9.5.

\begin{table*}
\centering
\caption{\label{tab:LTgrid} Spectral type definitions.}
\begin{tabular}{ccccccccccccccc}
Lum.~class:&I&I&I/III&III&III/V&V&V&I&I&I/III&III&III/V&V&V\\
Sp.~type&$L_1$&$L_2$&$L_3$&$L_4$&$L_5$&$L_6$&$L_7$&\Teff$_1$&\Teff$_2$&\Teff$_3$&\Teff$_4$&\Teff$_5$&\Teff$_6$&\Teff$_7$\\
\hline
\multicolumn{15}{c}{Milky Way, $z$ = 0.02}\\ 
\hline
2.0 &6.45 &6.08 &6.07 &6.06 &6.06 &6.05 &5.93 &43182.0 &45070.5 &45277.5 &45484.5 &46482.0 &47479.5 &50906.1\\
2.0/2.5&6.43&6.05&6.03&6.01&5.99&5.97&5.84&42128.0&44063.0&44265.2&44467.5&45479.0&46490.5&50119.0\\
2.5/3.0&6.41&6.02&5.98&5.95&5.92&5.88&5.71&41198.0&43055.0&43252.8&43450.5&44378.2&45306.0&48459.0\\
3.0/3.5&6.38&5.99&5.94&5.90&5.85&5.80&5.61&40221.0&42089.0&42333.5&42578.0&43447.5&44317.0&47698.0\\
3.5/4.0&6.35&5.96&5.90&5.85&5.79&5.72&5.50&38993.0&41164.5&41507.2&41850.0&42784.2&43718.5&46767.0\\
4.0/4.5&6.33&5.93&5.86&5.79&5.72&5.64&5.40&37883.0&40156.5&40574.0&40991.5&41970.5&42949.5&45987.0\\
4.5/5.0&6.29&5.89&5.82&5.73&5.65&5.56&5.31&36501.0&39065.5&39533.8&40002.0&41006.0&42010.0&44930.0\\
5.0/5.5&6.26&5.85&5.77&5.67&5.58&5.46&5.17&35426.0&37795.0&38275.0&38755.0&39778.0&40801.0&43384.0\\
5.5/6.0&6.23&5.80&5.71&5.60&5.49&5.36&5.02&34274.0&36408.5&36873.2&37338.0&38222.2&39106.5&41418.0\\
6.0/6.5&6.20&5.76&5.66&5.53&5.41&5.25&4.85&33161.0&35200.5&35684.5&36168.5&36828.5&37488.5&39197.0\\
6.5/7.0&6.16&5.72&5.61&5.46&5.33&5.15&4.68&31782.0&33990.0&34570.5&35151.0&35664.8&36178.5&37546.0\\
7.0/7.5&6.10&5.67&5.55&5.40&5.26&5.05&4.56&30118.0&32619.5&33341.0&34062.5&34518.8&34975.0&36067.0\\
7.5/8.0&6.07&5.62&5.50&5.33&5.18&4.95&4.45&29164.0&31461.0&32245.5&33030.0&33465.5&33901.0&34962.0\\
8.0/8.5&6.05&5.59&5.46&5.27&5.11&4.86&4.36&28443.0&30756.5&31443.8&32131.0&32541.8&32952.5&33796.0\\
8.5/9.0&6.03&5.56&5.42&5.21&5.04&4.77&4.26&27831.0&30036.5&30624.8&31213.0&31618.0&32023.0&32797.0\\
9.0/9.5&5.99&5.52&5.37&5.15&4.97&4.67&4.14&26781.0&28999.5&29741.8&30484.0&30745.0&31006.0&31476.0\\
9.5/B0.0&5.93&5.46&5.32&5.10&4.92&4.60&4.03&25003.0&27570.0&28610.2&29650.5&29872.2&30094.0&30458.0\\
\hline
\hline
\multicolumn{15}{c}{LMC, $z$ = 0.008}\\ 
\hline
2.0 &   6.41& 6.12 &6.05 &5.96 &5.91 &5.85 &5.68 &41508.5 &46752.5 &48104.8 &49457.0 &50667.8 &51878.5 &55897.2\\
2.0/2.5&6.38& 6.07 &6.00 &5.91 &5.85 &5.78 &5.61 &40283.8 &45653.5 &46947.2 &48241.0 &49450.2 &50659.5 &54943.8\\
2.5/3.0&6.37& 6.03 &5.95 &5.85 &5.79 &5.72 &5.50 &39683.0 &44555.0 &45790.0 &47025.0 &48233.0 &49441.0 &53444.5\\
3.0/3.5&6.35& 5.99 &5.90 &5.79 &5.73 &5.66 &5.39 &39022.4 &43457.0 &44633.0 &45809.0 &47016.0 &48223.0 &52034.5\\
3.5/4.0&6.33& 5.94 &5.85 &5.73 &5.66 &5.59 &5.30 &37894.6 &42358.5 &43475.8 &44593.0 &45798.8 &47004.5 &50877.5\\
4.0/4.5&6.31& 5.90 &5.80 &5.67 &5.60 &5.53 &5.18 &37213.6 &41259.5 &42318.2 &43377.0 &44581.2 &45785.5 &49402.0\\
4.5/5.0&6.29& 5.86 &5.75 &5.61 &5.54 &5.46 &5.07 &36491.9 &40161.0 &41161.0 &42161.0 &43364.0 &44567.0 &47899.7\\
5.0/5.5&6.26& 5.81 &5.70 &5.55 &5.48 &5.40 &4.98 &35433.6 &39063.0 &40004.0 &40945.0 &42147.0 &43349.0 &46752.2\\
5.5/6.0&6.24& 5.77 &5.65 &5.49 &5.42 &5.33 &4.86 &34696.0 &37964.5 &38846.8 &39729.0 &40929.8 &42130.5 &45189.8\\
6.0/6.5&6.22& 5.73 &5.60 &5.43 &5.36 &5.27 &4.74 &33924.5 &36865.5 &37689.2 &38513.0 &39712.2 &40911.5 &43611.8\\
6.5/7.0&6.19& 5.68 &5.55 &5.37 &5.30 &5.21 &4.64 &32916.7 &35767.0 &36532.0 &37297.0 &38495.0 &39693.0 &42351.4\\
7.0/7.5&6.16& 5.64 &5.51 &5.31 &5.24 &5.14 &4.55 &31972.0 &34669.0 &35375.0 &36081.0 &37278.0 &38475.0 &41106.1\\
7.5/8.0&6.14& 5.60 &5.46 &5.25 &5.17 &5.08 &4.42 &31193.2 &33570.5 &34217.8 &34865.0 &36060.8 &37256.5 &39428.9\\
8.0/8.5&6.11& 5.55 &5.41 &5.20 &5.11 &5.01 &4.31 &30225.8 &32471.5 &33060.2 &33649.0 &34843.2 &36037.5 &38048.1\\
8.5/9.0&6.08& 5.51 &5.36 &5.14 &5.05 &4.95 &4.19 &29428.4 &31373.0 &31903.0 &32433.0 &33626.0 &34819.0 &36405.0\\
9.0/9.5&6.06& 5.47 &5.31 &5.08 &4.99 &4.88 &4.05 &28610.5 &30275.0 &30746.0 &31217.0 &32409.0 &33601.0 &34681.6\\
9.5/B0.0&6.02& 5.42& 5.26 &5.02 &4.93 &4.82 &3.94 &27671.5 &29176.5 &29588.8 &30001.0 &31191.8 &32382.5 &33242.4\\
\hline
\hline
\multicolumn{15}{c}{SMC, $z$ = 0.004}\\ 
\hline
2.0& 6.53& 6.10& 6.07& 6.03& 5.86& 5.60& 4.96& 46917.2& 44750.0& 44602.5& 44455.0& 46227.0& 47999.0& 52669.3\\
2.0/2.5& 6.49& 6.06& 6.02& 5.97& 5.81& 5.55& 4.90& 45049.5& 43822.0& 43709.5& 43597.0& 45275.0& 46953.0& 51428.8\\
2.5/3.0& 6.46& 6.01& 5.97& 5.92& 5.76& 5.51& 4.85& 43774.8& 42893.5& 42816.5& 42739.5& 44323.2& 45907.0& 50368.6\\
3.0/3.5& 6.43& 5.96& 5.92& 5.87& 5.71& 5.46& 4.79& 42452.5& 41964.5& 41923.5& 41882.5& 43371.8& 44861.0& 49111.0\\
3.5/4.0& 6.40& 5.92& 5.87& 5.81& 5.66& 5.42& 4.74& 41089.1& 41036.0& 41030.5& 41025.0& 42420.0& 43815.0& 48033.9\\
4.0/4.5& 6.37& 5.87& 5.82& 5.76& 5.61& 5.38& 4.69& 39810.7& 40108.0& 40137.5& 40167.0& 41468.0& 42769.0& 46941.5\\
4.5/5.0& 6.34& 5.83& 5.77& 5.71& 5.56& 5.33& 4.63& 38628.1& 39179.5& 39244.5& 39309.5& 40516.2& 41723.0& 45651.1\\
5.0/5.5& 6.31& 5.78& 5.72& 5.65& 5.51& 5.29& 4.57& 37368.2& 38250.5& 38351.5& 38452.5& 39564.8& 40677.0& 44529.7\\
5.5/6.0& 6.28& 5.74& 5.67& 5.60& 5.46& 5.24& 4.51& 36280.7& 37322.0& 37458.5& 37595.0& 38613.0& 39631.0& 43206.9\\
6.0/6.5& 6.25& 5.69& 5.62& 5.54& 5.41& 5.20& 4.45& 35050.1& 36394.0& 36565.5& 36737.0& 37661.0& 38585.0& 42054.0\\
6.5/7.0& 6.22& 5.65& 5.58& 5.49& 5.35& 5.16& 4.41& 33833.2& 35465.5& 35672.5& 35879.5& 36709.2& 37539.0& 41021.0\\
7.0/7.5& 6.18& 5.60& 5.53& 5.44& 5.30& 5.11& 4.34& 32587.5& 34536.5& 34779.5& 35022.5& 35757.8& 36493.0& 39633.7\\
7.5/8.0& 6.15& 5.56& 5.48& 5.38& 5.25& 5.07& 4.28& 31619.5& 33608.0& 33886.5& 34165.0& 34806.0& 35447.0& 38400.6\\
8.0/8.5& 6.11& 5.51& 5.43& 5.33& 5.20& 5.02& 4.21& 30408.0& 32680.0& 32993.5& 33307.0& 33854.0& 34401.0& 36963.8\\
8.5/9.0& 6.08& 5.47& 5.38& 5.27& 5.15& 4.98& 4.15& 29471.1& 31751.5& 32100.5& 32449.5& 32902.2& 33355.0& 35654.6\\
9.0/9.5& 6.04& 5.42& 5.33& 5.22& 5.10& 4.94& 4.09& 28277.4& 30822.5& 31207.5& 31592.5& 31950.8& 32309.0& 34292.1\\
9.5/B0.0& 6.01& 5.38& 5.28& 5.17& 5.05& 4.89& 4.01& 27363.6& 29894.0& 30314.5& 30735.0& 30999.0& 31263.0& 32750.0\\
\hline
\end{tabular}
\end{table*}

\begin{figure}
\begin{center}
\includegraphics[width=8cm]{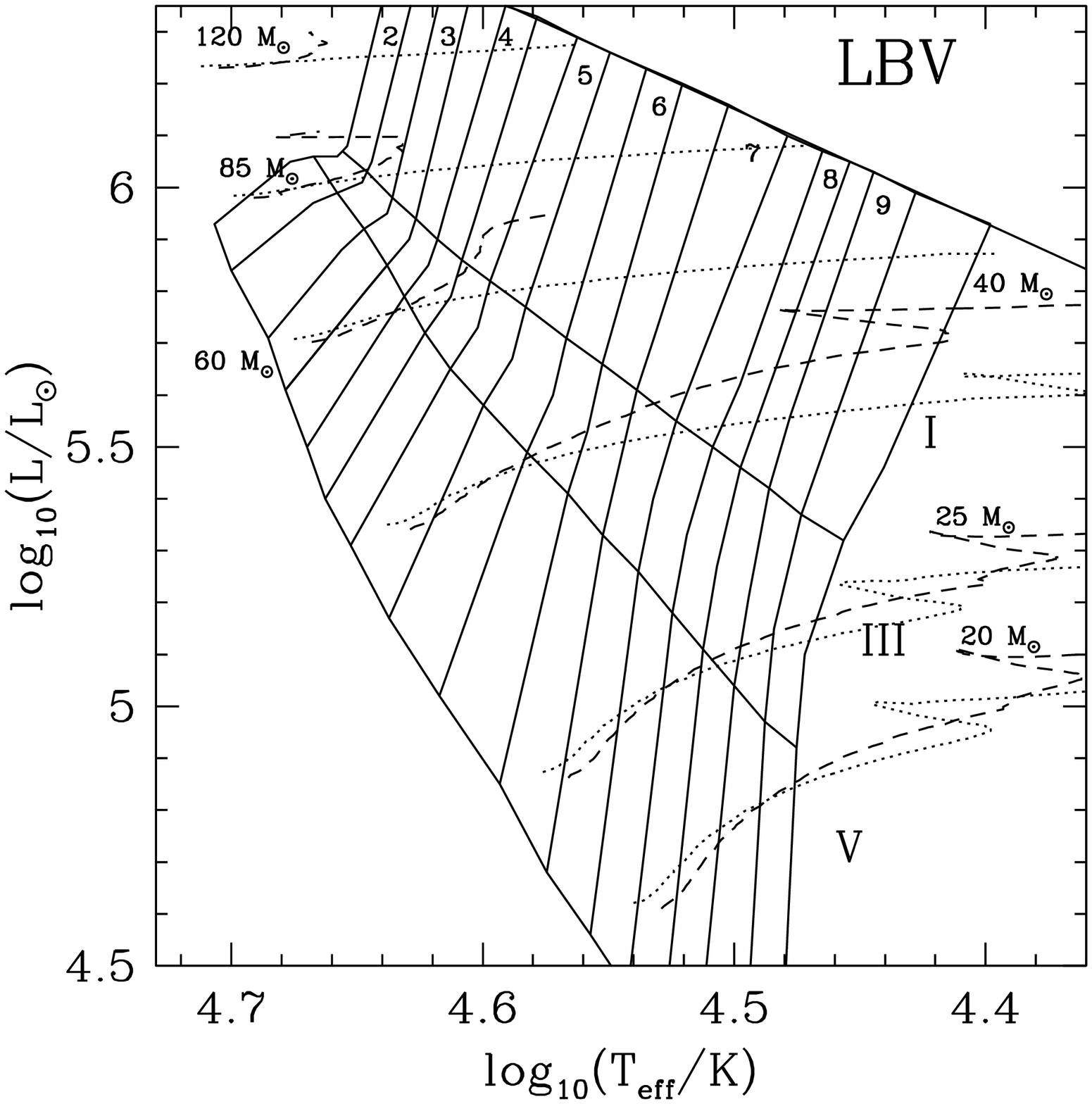}
\vspace*{-2.0cm}
\caption{Similar as {\it panel A} of Fig.~\ref{fig:sptypes}, but only
  the spectral subtype boxes are shown as a grid of {\it solid
  lines}. The {\it dashed lines} are rotating and the {\it dotted
  lines} non-rotating solar metallicity stellar evolutionary tracks by
  \citet{MM03} with initial masses as indicated in the figure. Only
  evolutionary phases before the Wolf-Rayet stage are plotted.}
\label{fig:tracks}
\end{center}
\end{figure}

In order to assign a certain spectral subtype and luminosity class to
a specific evolutionary point in time, stellar evolution models from
\citet{MM03} (see also appendix~\ref{app:evol}) are followed throughout 
the $T_\mathrm{eff}$-$L$-diagram. This is visualized in
Fig.~\ref{fig:tracks}, where the luminosity- and
$T_\mathrm{eff}$-evolution for six rotating as well as non-rotating
stellar models from 20 to 120 $M_\odot$ \citep{MM03} are plotted
(rotating models as {\it dashed lines} and non-rotating ones as {\it
  dotted lines}). In this figure, only the part of the evolution
before the Wolf-Rayet (WR) stage is depicted.
Objects that are still assumed to be core hydrogen are classified as WNL stars
\citep{HGL06}. However, we note that the WNL classification is only used in
appendix~\ref{app:stevol}. The present-day mass of a model
during its evolution through a spectral subclass is from now on 
referred to as evolutionary mass ($M_\mathrm{evol}$).

Table~\ref{tab:Orot} shows the new spectral type mass conversion, based
on solar metallicity rotating evolutionary models from 10 to 120
$M_\odot$ \citep{MM03}. The rotating models have initial rotational
velocities ($v_\mathrm{rot~ini}$) of 300 km/s, which results in 
$v_\mathrm{rot}$ during the Main-Sequence evolution of 180 to 240
km/s. These velocities are within the range observed for O stars
\citep{MKE06}. A table with non-rotating models is provided as
Table~\ref{tab:Onorot}. As the \citet{MM03} models only provide a
limited mass resolution, a special interpolation routine (described in
detail in appendix~\ref{app:evol}) is deployed in order to provide a
mass resolution down to 1 $M_\odot$. For LMC metallicity
\citet{MM05} provide only four models (30, 40, 60 and 120 $M_\odot$,
all rotating with 300 km/s), two additional models (15 and 20
$M_\odot$) are taken from the Padova group \citep{BNG09}. The resulting
spectral type mass conversion for LMC metallicity stars is shown in
Table~\ref{tab:Oz08}. For SMC metallicities, \citet{MM03}
only include three (all rotating) models (40, 60 and 120
$M_\odot$). Again, two models are added here (15 and 20 $M_\odot$)
from \citet{BNG09} in order to derive a spectral type mass conversion
(Table~\ref{tab:Oz04}).

The masses shown in the Tables~\ref{tab:Orot} to \ref{tab:Oz04} are all
weighted by the duration of the models in each spectral class. The
errors are assigned by using the most- and 
least-massive model entering the spectral class. As mentioned before,
each spectral class has an assumed error in \Teff~of 1000 K.
The minimal and maximal start and end ages give the range of possible
ages for the stars in a spectral-class box.
The advantage of using this method is the consistent application of
observational constraints for the different evolutionary phases on one
set of stellar evolution models. This allows one to place more
constraints for the stars in a certain spectral class on the range of
their initial and present-day masses.   

A number of other mass estimates for spectral types exist in the
literature, e.g.~\citet{VGS96} and \citet{HHC97}, but only the most
recent one by MSH05 is used here.
These models provide spectroscopic stellar masses that are derived
from the stellar luminosity, $L$, and \Teff~of NLTE stellar atmosphere
models through
\begin{equation}
\label{eq:M_spec}
M = \frac{gR^2}{G},
\end{equation}
where $G$ equals Newton's gravitational constant, and 
$g$ is the gravitational acceleration of the star at radius 
$R$:
\begin{equation}
\label{eq:M_R}
R = \sqrt{\frac{L}{4\pi\sigma_\mathrm{R}T_\mathrm{eff}^4}},
\end{equation}
where $\sigma_\mathrm{R}$ is the Stefan-Boltzmann constant.
\citet{MSH05} provide two mass estimates, one for a theoretical
\Teff~calibration ($m_\mathrm{MSH1}$) and one for an observational
\Teff~calibration ($m_\mathrm{MSH2}$).

Instead of using spectral type calibrations, a more direct way 
to derive spectroscopic masses is by carefully fitting model atmospheres to
high-resolution spectra, where both $g$, and \Teff~are determined
simultaneously \citep[see for example][]{RPH04}. Together with its
absolute magnitude it is possible to arrive at a mass using
Eqs.~(\ref{eq:M_spec}) and (\ref{eq:M_R}). A sample of spectroscopic
masses derived with this method will also be compared with dynamical
and model masses.

\subsection{Limitations}
Although the results presented here cover a large parameter space, they
involve some caveats. First of all, both the employed \citet{MM03} stellar
evolution models, as well as the MSH05 stellar atmospheres, which
define our spectral classes, only cover solar metallicity
(z~=~0.02). Metallicity is known to have a very strong influence
on the evolution and atmospheres of (O) stars via their
metallicity-dependent winds. The newly developed spectral type
definitions for LMC and SMC metallicities are a first step to loosen
these limitations, but are not as thoroughly based as the MSH05 work
for solar metallicity.

Table~\ref{tab:MM03z20} shows the spectral evolution of a series of
massive stellar models of different metallicities \citep[z~=~0.004,
  0.008, 0.02 and 0.04,][]{MM03,MM05} using the spectral type
definitions in Table~\ref{tab:LTgrid}. While these spectral type
definitions are based on solar metallicity atmospheres or empirical
\Teff~calibrations (for z=~0.008 and z=~0.004), considerable
differences in the evolution are noticeable.

Another relevant aspect for the evolution of massive
stars concerns binary evolution. Because many (if not most) massive stars
are part of a binary system, often with considerable secondary 
masses \citep{PBH99,ABK07,KF07,RCN09,WK07c,SMG10}, they could be capable of
influencing each other's evolution in a profound manner. 
Because all observations presented in Table~\ref{tab:dyn}
involve eclipsing binaries, all the objects must form tight pairs with
reasonably large stars, and therefore binary evolution is bound to be
important, but it is a non-trivial matter to account for it.

As was mentioned in the introduction, an additional potential prime source for 
errors in the mass determination concerns the atmosphere and wind
parameters, as well as the mass-loss prescription employed in the
evolutionary models.

\section{Dynamical masses of eclipsing binaries}
\label{sec:data}

In recent years, observational techniques allowed us to measure masses
of very massive stars directly by observing the orbits of massive
eclipsing binaries. In Table~\ref{tab:dyn} the dynamical mass
estimates for 33 very massive stars are listed. The majority of the
stars (22) are from a compilation by \citet{G03}, who provides three
lists with massive binaries from the literature. His first list
shows detached systems, the second one non-eclipsing binaries (with
lower mass limits only) and the third systems, which are either
dynamically evolved (semi-detached or contact systems) or contain
giants or supergiants. All but two systems from the first list are included
in Table~\ref{tab:dyn} as are three systems form the third list, two of
them are given as being before the interaction stage and the
supergiant V729 Cyg. The systems with lower limits only and the ones
which are dynamically evolved are not suitable for the current study
and are therefore not included. The remaining eclipsing binaries
except one are from literature published after the \citet{G03} list,
but which contain the necessary data for this study. The exception is 
WR22 B, which is not covered in \citet{G03} because the primary is a
Wolf-Rayet star.

The dynamical masses from Table~\ref{tab:dyn} involve 
present-day masses instead of initial masses. This is accounted for 
by not only comparing the evolving parameters with the luminosity and
\Teff~grid, but by simultaneously keeping track of the initial
mass. Therefore, in Table~\ref{tab:Orot} 
the initial stellar mass for a spectral type is given as well as the
possible minimal and maximal mass when the stars enter and leave the
respective spectral type. Additionally, the minimal and maximal age is
given when the models enter and leave a spectral type.

\begin{table*}
\centering
\caption{\label{tab:dyn} Eclipsing O-star binaries with dynamical mass
  estimates.}
\begin{tabular}{cccccccccl}
Star&Sp Type&$m_{\rm dynamical}$&$m_{\rm MSH 1}$&$m_\mathrm{MSH 2}$&$m_\mathrm{ini}$&$m_\mathrm{evol}$&$m_\mathrm{start}$&$m_\mathrm{end}$&Ref.\\ 
\hline
\multicolumn{10}{c}{MW}\\ 
\hline
 HD93205A             & O3V    &  56.0$\pm$   4.0&  58.3&  58.0&  67 -9/+9&   65 -7/+7&   67 -9/+6&   64 -6/+8& (1)\\
 FO15 A               & O5.5V  &  30.0$\pm$   1.0&  34.2&  34.4&  40 -7/+8&   39 -6/+7&   39 -6/+6&   38 -5/+7& (2)\\
 FO15 B               & O9.5V  &  16.0$\pm$   1.0&  16.5&  15.6&  18 -5/+4&   18 -5/+4&   18 -5/+4&   18 -5/+4& (2)\\
 Theta Orionis C1     & O6Vpe  &  35.8$\pm$   7.2&  31.7&  31.0&  35 -7/+6&   34 -6/+5&   34 -6/+5&   33 -6/+6& (3)\\
 V1036 Sco A          & O6V    &  32.0$\pm$   4.0&  31.7&  31.0&  35 -7/+6&   34 -6/+5&   34 -6/+5&   33 -6/+6& (4)\\
 V1036 Sco B          & O7V    &  32.0$\pm$   4.0&  26.5&  25.3&  28 -6/+8&   27 -5/+7&   27 -5/+6&   27 -5/+7& (4)\\
 LS1135 A             & O6.5V  &  30.0$\pm$   1.0&  29.0&  28.0&  31 -7/+6&   30 -6/+6&   30 -6/+5&   30 -6/+6& (5)\\
 V729 Cyg             & O7Ianfp&  47.0$\pm$   9.0&  40.9&  38.4&  47 -9/+36&  39 -5/+23&  40 -6/+23&  39 -5/+24& (4)\\
 V1007 Sco A          & O7.5III\tablefootmark{a}&  29.5$\pm$   0.4&  29.1&  27.4&  33 -5/+6&   31 -4/+4&   31 -5/+4&   31 -4/+5& (6)\\
 V1007 Sco B          & O7III\tablefootmark{a}  &  30.1$\pm$   0.4&  31.2&  29.6&  36 -5/+7&   34 -4/+5&   34 -4/+5&   33 -4/+5& (6)\\
 V3903 Sgr A          & O7V    &  27.3$\pm$   6.0&  26.5&  25.3&  28 -6/+8&   27 -5/+7&   27 -5/+6&   27 -5/+7& (4)\\
 V3903 Sgr B          & O9V    &  19.0$\pm$   4.0&  18.0&  17.1&  19 -5/+5&   19 -5/+4&   19 -5/+4&   19 -5/+4& (4)\\
 CPD -59 2603 A       & O7V    &  22.7$\pm$ 4.0\tablefootmark{b}&  26.5&  25.3&  28 -6/+8&   27 -5/+7&   27 -5/+6&   27 -5/+7& (4)\\
 CPD -59 2603 B       & O9.5V  &  14.5$\pm$ 4.0\tablefootmark{b}&  16.5&  15.6&  18 -5/+4&   18 -5/+4&   18 -5/+4&   18 -5/+4& (4)\\
 V1182 Aql A          & O8Vnn  &  31.0$\pm$   0.6&  21.9&  20.8&  23 -6/+6&   22 -5/+6&   22 -5/+6&   22 -5/+6& (7)\\
 EM Car A             & O8V    &  22.9$\pm$   3.0&  21.9&  20.8&  23 -6/+6&   22 -5/+6&   22 -5/+6&   22 -5/+6& (4)\\
 EM Car B             & O8V    &  21.4$\pm$   3.0&  21.9&  20.8&  23 -6/+6&   22 -5/+6&   22 -5/+6&   22 -5/+6& (4)\\
 CC Cas               & O8.5III&  18.3$\pm$   5.0&  24.8&  23.7&  29 -5/+6&   27 -4/+5&   27 -4/+5&   27 -4/+5& (4)    \\
 WR22 B               & O9V    &  20.6$\pm$   1.7&  18.0&  17.1&  19 -5/+5&   19 -5/+4&   19 -5/+4&   19 -5/+4& (8)\\
 V478 Cyg A           & O9.5V  &  16.6$\pm$   9.0&  16.5&  15.6&  18 -5/+4&   18 -5/+4&   18 -5/+4&   18 -5/+4& (4)\\
 V478 Cyg B           & O9.5V  &  16.3$\pm$   9.0&  16.5&  15.6&  18 -5/+4&   18 -5/+4&   18 -5/+4&   18 -5/+4& (4)\\
 CPD -59 2628 A       & O9.5V  &  14.0$\pm$  20.0&  16.5&  15.6&  18 -5/+4&   18 -5/+4&   18 -5/+4&   18 -5/+4& (4)\\
\hline
\hline
\multicolumn{10}{c}{LMC}\\ 
\hline
 LMC MACHO 053441.3 A & O3If   &  41.2$\pm$   12.0&  66.9&  67.5&  81 -10/+9& 72 -8/+9& 82 -10/+6& 76 -5/+7& (4)    \\
 LMC MACHO 053441.3 B & O6V    &  27.0$\pm$   12.0&  31.7&  31.0&  37 -5/+3& 36 -4/+3& 37 -5/+3& 36 -4/+3& (4)    \\
 LMC R136-38 A        & O3V    &  56.9$\pm$   6.0&  58.3&  58.0&  64 -7/+7& 64 -7/+7& 64 -7/+7& 64 -7/+7& (4)\\
 LMC R136-38 B        & O6V    &  23.4$\pm$   2.0&  31.7&  31.0&37 -5/+3& 36 -4/+3& 37 -5/+3& 36 -4/+3& (4)    \\
 LMC R136-42 A        & O3V    &  40.3$\pm$   1.0&  58.3&  58.0&  64 -7/+7& 64 -7/+7& 64 -7/+7& 64 -7/+7& (4)\\
 LMC R136-42 B        & O3V    &  32.6$\pm$   1.0&  58.3&  58.0&  64 -7/+7& 64 -7/+7& 64 -7/+7& 64 -7/+7& (4)\\
 LH 54-425 A          & O3V    &  50.0$\pm$  10.0&  58.3&  58.0&  64 -7/+7& 64 -7/+7& 64 -7/+7& 64 -7/+7& (9)\\
 LH 54-425 B          & O5V    &  30.0$\pm$   6.0&  37.3&  38.1&  44 -5/+5& 44 -5/+5& 44 -5/+4& 43 -4/+5& (9)\\
 LMC R136-77 A        & O5.5V  &  28.9$\pm$   3.0&  34.2&  34.4&  40 -5/+4& 40 -5/+4& 40 -5/+4& 39 -4/+4& (4)\\
 LMC R136-77 B        & O5.5V  &  26.2$\pm$   3.0&  34.2&  34.4&  40 -5/+4& 40 -5/+4& 40 -5/+4& 39 -4/+4& (4)\\
 LMC-SC1-105 A        & O7V    &  30.9$\pm$   1.0&  26.5&  25.3&  31 -5/+3& 30 -4/+3& 31 -5/+3& 30 -4/+3& (10)\\
\hline
\end{tabular}
\tablefoot{For these massive stars
  dynamical mass estimates, $m_\mathrm{dyn}$, exist from the orbits of
  binaries. The other mass estimates are from the theoretical
  \Teff~calibration ($m_{\rm MSH 1}$) and observational
  \Teff~calibration ($m_\mathrm{MSH 2}$) by MSH05, together with the
  initial ($m_\mathrm{ini}$), mean evolutionary ($m_\mathrm{evol}$), minimal
  ($m_\mathrm{start}$) and maximal ($m_\mathrm{end}$) present-day mass 
from this work. The horizontal lines separates objects in the
LMC from Galactic ones. All masses are in $M_\odot$. For the Milky Way
stars Table~\ref{tab:Orot} was used to derive the masses and for the
LMC stars Table~\ref{tab:Oz08}.
\tablefoottext{a}{Different luminosity class determinations exist in the
literature. The most recent one was used.} 
\tablefoottext{b}{Error assumed as none given.}
\tablefoottext{c}{Luminosity class V was assumed.}
}
\tablebib{
1:~\citet{MBN01,G03},
2:~\citet{NMF06},
3:~\citet{KWB08},
4:~\citet{G03},
5:~\citet{FN06},
6:~\citet{MHN08},
7:~\citet{MDL05b},
8:~\citet{SSS99},
9:~\citet{WGH08},
10:~\citet{B09}}

\end{table*}

In addition to dynamical mass determinations, several spectroscopic masses 
exist for O-type stars. Table~\ref{tab:spec} shows a compilation
of these spectroscopic masses taken from \citet{RPH04}, besides the 
MSH05 masses and the evolutionary masses presented in this work.

\begin{table*}
\centering
\caption{\label{tab:spec} O-star with spectroscopic mass estimates.}
\begin{tabular}{ccccccc}
Star&Sp Type&$m_\mathrm{spectroscopic}$&$m_\mathrm{MSH1}$&$m_\mathrm{MSH2}$&$m_\mathrm{evol}$\\
\hline
HD93129A\tablefootmark{a}&O2If$^\ast$& 94.8 -28.8/+41.3&66.9$^+$&67.5$^+$&95
-33/+25\\
HD14947& O5I& 30.7 -9.2/+13.1 &50.9 &50.7&47 -4/+17\\
HD210839\tablefootmark{b}& O6I& 62.2 -24.9/+41.5 &45.8 &44.1&42 -3/+15\\
HD192639& O7I& 37.5 -11.2/+16.1 &40.9 &38.4&39 -5/+24\\
HD193514& O7I& 28.2 -8.5/+12.1 &40.9 &38.4&39 -5/+24\\
HD210809& O9I& 21.7 -6.6/+9.4 &32.0 &29.6&33 -5/+11\\
HD207198\tablefootmark{a}& O9I& 29.0 -8.7/+12.5 &32.0 &29.6&33 -5/+11\\
HD30614& O9.5I& 37.6 -11.2/+16.1 &30.4 &27.8&32 -5/+13\\
HD209975\tablefootmark{a}& O9.5I& 31.4 -9.4/+13.4 &30.4 &27.8&32 -5/+13\\
HD15558\tablefootmark{a}&O5III& 78.7 -23.7/+33.8 &41.5 &40.4&47 -5/+6\\
HD193682&O5III& 27.9 -8.2/+11.7 &41.5 &40.4&47 -5/+6\\
HD190864&O6.5III& 20.3 -6.1/+8.7 &33.7 &32.0&36 -4/+5\\
HD24912&O7.5III& 26.1 -7.6/+10.9 &29.1 &27.5&31 -4/+5\\
HD203064&O7.5III&35.9 10.3/+14.9 &29.1 &27.5&31 -4/+5\\
HD191423\tablefootmark{c}&O9III& 24.6 -7.0/+11.2 &23.1 &22.0&26 -4/+5\\
HD93128&O3V& 39.8 -12.0/+17.2 &58.3 &58.0&65 -7/+8\\
HD93250&O3V& 83.3 -25.1/+36.0 &58.3 &58.0&65 -7/+8\\
HD66811&O4V& 53.9 -19.5/+30.8 &46.2 &46.9&54 -6/+7\\
HD15629\tablefootmark{a}&O5V& 30.4 -9.1/+13.1 &37.3 &38.1&44 -6/+7\\
HD217086&O7V& 14.2 -4.0/+6.3 &26.5 &25.3&27 -5/+7\\
HD149757\tablefootmark{b}&O9V& 20.2 -5.7/+8.8 &18.0 &17.1&19 -5/+5\\
\hline
\end{tabular}
\tablefoot{Mass estimates arrived at by spectral line
  fitting ($m_\mathrm{spectroscopic}$) from \citet{RPH04}. Additionally, the mass
  estimates from MSH05 ($m_\mathrm{MSH1}$ and $m_\mathrm{MSH2}$) and
  this work ($m_\mathrm{evol}$) are shown. All masses are in $M_\odot$.
\tablefoottext{a}{Member of a binary system.} 
\tablefoottext{b}{Runaway star.}
\tablefoottext{c}{Extremely fast rotator.}
}
\end{table*}

\section{Results and discussion}
\label{sec:disc}

The Tables~\ref{tab:Orot} and \ref{tab:Onorot} show the determined
initial masses as well as the mean, minimal, and maximal present-day
masses according to the \citet{MM03} rotating (300 km/s) and
non-rotating stellar evolution models, and the MSH05 O star spectral
type definition. Tables~\ref{tab:Oz08} and \ref{tab:Oz04} show
the same for LMC and SMC metallicities, respectively. The errors
shown for the mass determination are the 
lowest mass and maximum mass models that pass through the spectral
type. They also include an error margin of $\pm$ 1000 K for the MSH05
spectral type definitions. The differences in the supergiant mass
errors from one subtype to the next have two main reasons. Spectral
types later than O 6.5 I are reached during stellar evolution from the
hot end by more massive stars and the cold end by less massive
stars. This results in a larger range of possible
masses. Also, the subtypes are not of the same area in the
$L$-\Teff~-space (see Fig.~\ref{fig:sptypes}). Therefore, some subtypes
simply have a higher probability to be encountered by the model tracks.
It would be possible to reduce these errors by introducing more
luminosity classes, like II, Ia and Ib. But no MSH05 definitions for
these classes presently exist.

Furthermore, the table shows the mean time the models
spend in each spectral type box. Again, the errors are defined by the
lowest and most-massive model passing through the spectral type box, 
including a $\pm$ 1000 K uncertainty for the spectral type definitions.
Note that the \citet{MM03} models show considerable jumps in \Teff~and
luminosity when the stars enter the Wolf-Rayet phase. These extremely
fast crossings ($<$ 50000 years) through the HR-Diagram are not
included in the tables.

\begin{table*}
\centering
\caption{\label{tab:Orot} Theoretical masses for O stars from rotating
  stellar models.}
\begin{tabular}{ccccccccc}
Spectral&$m_\mathrm{ini}$&$m_\mathrm{evol}$&$m_\mathrm{start}$&$m_\mathrm{end}$&
$t_\mathrm{start, min}$& $t_\mathrm{start, max}$& $t_\mathrm{end,
  min}$& $t_\mathrm{end, max}$\\
type&$M_\odot$&$M_\odot$&$M_\odot$&$M_\odot$&Myr&Myr&Myr&Myr\\
\hline
O 2.0 If$^\ast$&107 -27/+13& 95 -33/+25&107 -45/+13& 83 -21/+11&0.0&1.5&0.1&1.8\\
O 2.0 I& 95 -18/+22& 79 -19/+15& 84 -24/+10& 75 -14/+11&0.9&1.8&1.2&2.0\\
O 2.5 I& 86 -14/+27& 72 -14/+14& 76 -18/+9& 67 -9/+16&1.2&2.1&1.4&2.3\\
O 3.0 I& 79 -10/+29& 66 -10/+16& 69 -12/+12& 62 -7/+18&1.4&2.5&1.6&2.5\\
O 3.5 I& 72 -8/+31& 60 -8/+18& 62 -10/+15& 58 -5/+20&1.6&2.4&1.9&2.5\\
O 4.0 I& 66 -7/+31& 55 -9/+19& 57 -7/+17& 54 -7/+21&1.9&2.8&2.2&2.8\\
O 4.5 I& 60 -4/+33& 50 -5/+23& 52 -6/+20& 47 -3/+24&2.2&2.9&2.5&3.1\\
O 5.0 I& 57 -5/+26& 47 -4/+17& 48 -5/+15& 45 -2/+18&2.6&3.2&2.9&3.4\\
O 5.5 I& 54 -6/+23& 44 -3/+16& 45 -4/+14& 43 -2/+17&2.9&3.9&3.2&4.0\\
O 6.0 I& 51 -6/+21& 42 -3/+15& 43 -4/+14& 41 -2/+16&3.3&4.1&3.6&4.2\\
O 6.5 I& 47 -6/+26& 39 -3/+18& 40 -4/+17& 39 -3/+18&3.6&4.3&3.9&4.5\\
O 7.0 I& 47 -9/+36& 39 -5/+24& 40 -6/+23& 39 -5/+24&3.8&4.7&4.1&4.7\\
O 7.5 I& 43 -7/+39& 37 -4/+26& 37 -5/+25& 36 -4/+26&4.2&5.1&4.5&5.1\\
O 8.0 I& 42 -7/+38& 35 -4/+26& 36 -5/+24& 35 -3/+26&4.6&5.3&4.7&5.4\\
O 8.5 I& 40 -7/+41& 33 -4/+25& 34 -4/+25& 33 -3/+11&4.7&5.7&4.8&5.7\\
O 9.0 I& 40 -9/+43& 33 -5/+11& 33 -5/+11& 33 -4/+11&4.9&6.0&4.9&6.0\\
O 9.5 I& 37 -8/+47& 32 -5/+13& 32 -5/+12& 31 -5/+13&5.0&6.4&5.1&6.4\\
\hline
O 2.0 III& 86 -12/+7& 77 -18/+12& 78 -19/+11& 76 -17/+11&0.6&1.4&0.7&1.8\\
O 2.5 III& 79 -10/+7& 71 -14/+9& 73 -16/+7& 69 -12/+11&0.8&1.8&1.0&2.1\\
O 3.0 III& 73 -8/+7& 65 -11/+8& 68 -13/+6& 62 -8/+11&1.0&2.1&1.2&2.5\\
O 3.5 III& 67 -7/+8& 60 -8/+8& 62 -7/+6& 58 -6/+10&1.1&2.8&1.3&2.8\\
O 4.0 III& 62 -6/+8& 56 -6/+8& 57 -5/+6& 54 -5/+8&1.2&2.7&1.4&3.0\\
O 4.5 III& 57 -6/+7& 51 -4/+7& 52 -5/+5& 50 -3/+7&1.4&2.9&1.7&3.2\\
O 5.0 III& 52 -6/+7& 47 -5/+6& 48 -6/+4& 46 -4/+6&1.6&3.2&2.0&3.6\\
O 5.5 III& 47 -7/+8& 43 -5/+6& 44 -6/+5& 42 -4/+6&1.9&3.5&2.4&3.9\\
O 6.0 III& 43 -6/+7& 39 -4/+6& 39 -5/+5& 39 -4/+5&2.4&3.8&2.8&4.1\\
O 6.5 III& 39 -5/+7& 36 -4/+5& 37 -4/+5& 36 -4/+5&2.8&4.1&3.4&4.3\\
O 7.0 III& 36 -5/+7& 34 -4/+6& 34 -4/+5& 33 -4/+5&3.2&4.6&3.7&4.7\\
O 7.5 III& 33 -5/+6& 31 -4/+5& 31 -5/+4& 31 -4/+5&3.7&5.0&4.1&5.2\\
O 8.0 III& 31 -6/+6& 29 -5/+5& 29 -5/+5& 29 -5/+5&4.1&5.5&4.5&5.6\\
O 8.5 III& 29 -5/+6& 27 -4/+5& 27 -4/+5& 27 -4/+5&4.5&5.9&4.9&6.1\\
O 9.0 III& 27 -5/+6& 26 -4/+5& 26 -4/+5& 26 -4/+5&5.1&6.6&5.3&6.7\\
O 9.5 III& 26 -5/+5& 24 -4/+5& 24 -4/+4& 24 -4/+4&5.5&7.1&5.7&7.3\\
\hline
O 2.0 V& 86 -11/+8& 83 -23/+9& 85 -25/+8& 81 -21/+11&0.0&1.1&0.1&1.3\\
O 2.5 V& 76 -12/+7& 73 -10/+7& 75 -11/+6& 71 -8/+8&0.0&1.0&0.1&1.5\\
O 3.0 V& 67 -9/+9& 65 -7/+8& 67 -9/+6& 64 -6/+9&0.0&1.4&0.1&1.8\\
O 3.5 V& 61 -8/+8& 59 -6/+7& 60 -7/+6& 58 -5/+8&0.0&1.8&0.1&1.9\\
O 4.0 V& 56 -8/+7& 54 -6/+7& 55 -7/+5& 53 -5/+7&0.0&1.7&0.1&2.1\\
O 4.5 V& 51 -8/+6& 49 -6/+6& 50 -7/+5& 49 -6/+6&0.0&2.0&0.1&2.4\\
O 5.0 V& 45 -7/+8& 44 -6/+7& 45 -7/+6& 43 -5/+7&0.0&2.2&0.1&2.7\\
O 5.5 V& 40 -7/+8& 39 -6/+7& 39 -6/+6& 38 -5/+7&0.0&2.6&0.1&3.1\\
O 6.0 V& 35 -7/+6& 34 -6/+6& 34 -6/+5& 33 -6/+6&0.0&3.1&0.2&3.6\\
O 6.5 V& 31 -7/+6& 30 -6/+6& 30 -6/+5& 30 -6/+6&0.0&3.6&0.1&4.1\\
O 7.0 V& 28 -6/+8& 27 -5/+7& 27 -5/+6& 27 -5/+7&0.0&4.1&0.4&4.6\\
O 7.5 V& 25 -5/+7& 25 -5/+7& 25 -5/+6& 25 -5/+6&0.0&4.6&0.3&5.1\\
O 8.0 V& 23 -6/+6& 22 -5/+6& 22 -5/+6& 22 -5/+6&0.0&5.1&0.1&5.5\\
O 8.5 V& 21 -6/+5& 21 -6/+5& 21 -6/+5& 21 -6/+5&0.0&5.5&0.2&6.1\\
O 9.0 V& 19 -5/+5& 19 -5/+5& 19 -5/+4& 19 -5/+4&0.0&6.2&0.2&6.8\\
O 9.5 V& 18 -5/+4& 18 -5/+4& 18 -5/+4& 18 -5/+4&0.0&6.8&0.2&7.5\\
\end{tabular}
\tablefoot{Theoretical masses for O stars from rotating
  solar metallicity stellar evolution models \citep{MM03}. The first mass
  ($m_\mathrm{ini}$) denotes the initial mass of the 
  model weighted by the time the star resides in that spectral
  type. The weighting is done in order to present the most likely mass
  for a spectral type. The lower and upper mass limit show which range
  of initial masses can reach a certain spectral type. The mean mass of
  the star which stays longest in a certain spectral type is denoted
  by $m_\mathrm{evol}$. The third mass
  ($m_\mathrm{start}$) is the mass with which the star starts
  when entering this spectral type while the fourth mass
  ($m_\mathrm{end}$) is the mass at the end of the stay in that
  particular spectral type. For each subclass is also given the
  minimal and maximal age the models when they enter it
  ($t_\mathrm{start, min}$, $t_\mathrm{start, max}$) and when the
  leave ($t_\mathrm{end, min}$, $t_\mathrm{end, max}$).}
\end{table*}

\begin{table*}
\centering
\caption{\label{tab:Onorot} Theoretical masses for O stars from non-rotating
  stellar models.}
\begin{tabular}{ccccccccc}
Spectral&$m_\mathrm{ini}$&$m_\mathrm{evol}$&$m_\mathrm{start}$&$m_\mathrm{end}$&
$t_\mathrm{start, min}$& $t_\mathrm{start, max}$& $t_\mathrm{end,
  min}$& $t_\mathrm{end, max}$\\
type&$M_\odot$&$M_\odot$&$M_\odot$&$M_\odot$&Myr&Myr&Myr&Myr\\
\hline
O 2.0 If$^\ast$&105 -23/+15&101 -19/+19&105 -23/+15& 96 -14/+13&0.0&1.0&0.1&1.2\\
O 2.0 I&103 -17/+17& 93 -15/+15& 94 -16/+14& 93 -15/+15&0.9&1.4&1.1&1.4\\
O 2.5 I& 99 -18/+21& 89 -16/+18& 90 -16/+17& 89 -15/+18&1.1&1.5&1.2&1.6\\
O 3.0 I& 97 -21/+23& 86 -18/+20& 87 -18/+19& 86 -17/+19&1.2&1.7&1.3&1.7\\
O 3.5 I& 96 -24/+24& 84 -19/+21& 84 -19/+20& 83 -19/+20&1.3&1.8&1.4&1.9\\
O 4.0 I& 94 -27/+26& 81 -21/+22& 82 -21/+21& 81 -21/+21&1.4&2.0&1.5&2.0\\
O 4.5 I& 91 -29/+29& 78 -22/+24& 79 -23/+23& 78 -22/+22&1.5&2.2&1.6&2.3\\
O 5.0 I& 78 -20/+42& 68 -16/+32& 68 -16/+31& 68 -15/+31&1.7&2.5&1.7&2.5\\
O 5.5 I& 68 -14/+49& 60 -11/+37& 60 -11/+36& 60 -11/+36&1.7&2.7&1.8&2.8\\
O 6.0 I& 66 -16/+19& 58 -13/+15& 58 -13/+14& 58 -12/+15&2.1&3.0&2.1&3.0\\
O 6.5 I& 64 -18/+21& 56 -14/+17& 56 -14/+16& 56 -14/+16&2.2&3.3&2.2&3.3\\
O 7.0 I& 61 -20/+24& 53 -16/+18& 54 -16/+18& 53 -15/+18&2.3&3.7&2.3&3.7\\
O 7.5 I& 53 -14/+32& 47 -11/+24& 47 -11/+24& 47 -11/+24&2.3&3.9&2.3&4.0\\
O 8.0 I& 48 -11/+35& 43 -9/+27& 43 -8/+26& 43 -9/+26&2.4&4.3&2.4&4.3\\
O 8.5 I& 46 -10/+35& 41 -8/+27& 42 -8/+26& 41 -8/+27&2.5&4.4&2.5&4.5\\
O 9.0 I& 45 -11/+15& 40 -9/+12& 41 -9/+11& 40 -8/+11&3.1&4.8&3.1&4.8\\
O 9.5 I& 44 -13/+16& 38 -8/+15& 40 -10/+12& 36 -7/+14&3.1&5.3&3.2&5.4\\
\hline
O 2.0 III& 88 -5/+6& 82 -6/+7& 83 -6/+6& 82 -5/+7&0.6&1.2&0.8&1.4\\
O 2.5 III& 81 -5/+6& 75 -5/+6& 76 -5/+5& 74 -4/+7&0.8&1.4&1.0&1.5\\
O 3.0 III& 74 -6/+7& 69 -6/+7& 70 -7/+6& 68 -5/+7&1.0&1.6&1.2&1.7\\
O 3.5 III& 69 -7/+7& 64 -6/+7& 64 -6/+6& 63 -6/+7&1.1&1.7&1.3&1.9\\
O 4.0 III& 64 -7/+8& 59 -6/+8& 59 -6/+7& 58 -5/+8&1.3&1.9&1.4&2.1\\
O 4.5 III& 59 -7/+8& 54 -6/+8& 55 -6/+7& 54 -5/+7&1.4&2.1&1.6&2.3\\
O 5.0 III& 54 -6/+9& 50 -5/+8& 51 -5/+7& 50 -5/+8&1.6&2.4&1.9&2.6\\
O 5.5 III& 49 -8/+9& 45 -7/+8& 46 -7/+7& 45 -6/+8&1.9&2.8&2.2&2.9\\
O 6.0 III& 44 -6/+10& 41 -5/+9& 42 -5/+8& 41 -5/+8&2.3&3.2&2.5&3.3\\
O 6.5 III& 41 -6/+9& 38 -5/+8& 39 -5/+7& 38 -5/+7&2.7&3.6&2.8&3.7\\
O 7.0 III& 38 -6/+8& 36 -5/+7& 36 -5/+7& 36 -5/+7&3.1&4.1&3.1&4.1\\
O 7.5 III& 34 -5/+8& 33 -5/+7& 33 -5/+6& 33 -5/+6&3.4&4.5&3.4&4.6\\
O 8.0 III& 32 -6/+7& 30 -5/+7& 30 -5/+6& 30 -5/+6&3.7&5.0&3.7&5.1\\
O 8.5 III& 30 -6/+7& 28 -5/+7& 28 -5/+6& 28 -5/+6&4.0&5.5&4.0&5.6\\
O 9.0 III& 28 -5/+7& 27 -4/+6& 27 -4/+6& 27 -4/+6&4.3&5.9&4.4&6.0\\
O 9.5 III& 27 -6/+6& 26 -5/+6& 26 -5/+5& 26 -5/+5&4.6&6.7&4.7&6.7\\
\hline
O 2.0 V& 85 -11/+8& 82 -9/+8& 84 -10/+7& 81 -7/+9&0.0&1.0&0.1&1.2\\
O 2.5 V& 74 -13/+9& 72 -11/+8& 74 -13/+6& 71 -10/+8&0.0&1.1&0.1&1.4\\
O 3.0 V& 66 -10/+10& 64 -8/+9& 65 -9/+7& 63 -7/+9&0.0&1.4&0.1&1.6\\
O 3.5 V& 60 -9/+9& 58 -7/+8& 59 -8/+7& 57 -6/+8&0.0&1.5&0.1&1.7\\
O 4.0 V& 54 -8/+9& 53 -7/+8& 53 -7/+7& 52 -6/+8&0.0&1.7&0.1&1.9\\
O 4.5 V& 49 -8/+8& 48 -7/+7& 49 -8/+6& 48 -7/+7&0.0&1.9&0.1&2.1\\
O 5.0 V& 44 -7/+9& 43 -6/+8& 44 -7/+7& 42 -6/+8&0.0&2.1&0.1&2.4\\
O 5.5 V& 39 -7/+9& 38 -6/+8& 39 -7/+7& 38 -6/+8&0.0&2.4&0.1&2.8\\
O 6.0 V& 34 -8/+8& 33 -7/+8& 33 -7/+7& 33 -7/+7&0.0&2.9&0.1&3.3\\
O 6.5 V& 30 -7/+8& 29 -6/+8& 29 -6/+7& 29 -6/+7&0.0&3.3&0.1&3.7\\
O 7.0 V& 27 -6/+8& 26 -5/+8& 26 -5/+7& 26 -5/+7&0.0&3.7&0.2&4.2\\
O 7.5 V& 24 -5/+8& 24 -5/+7& 24 -5/+7& 24 -5/+7&0.0&4.2&0.1&4.6\\
O 8.0 V& 22 -5/+7& 22 -5/+7& 22 -5/+6& 22 -5/+6&0.0&4.7&0.2&5.2\\
O 8.5 V& 20 -5/+6& 20 -5/+6& 20 -5/+5& 20 -5/+5&0.0&5.2&0.2&5.7\\
O 9.0 V& 19 -5/+5& 19 -5/+5& 19 -5/+5& 19 -5/+5&0.0&5.7&0.2&6.4\\
O 9.5 V& 17 -4/+5& 17 -4/+5& 17 -4/+4& 17 -4/+4&0.0&6.4&0.2&7.2\\
\hline
\end{tabular}
\end{table*}

\begin{table*}
\centering
\caption{\label{tab:Oz08} Theoretical masses for O stars from rotating
  LMC (z=0.008) stellar models.}
\begin{tabular}{ccccccccc}
Spectral&$m_\mathrm{ini}$&$m_\mathrm{evol}$&$m_\mathrm{start}$&$m_\mathrm{end}$&
$t_\mathrm{start, min}$& $t_\mathrm{start, max}$& $t_\mathrm{end,
  min}$& $t_\mathrm{end, max}$\\
type&$M_\odot$&$M_\odot$&$M_\odot$&$M_\odot$&Myr&Myr&Myr&Myr\\
\hline
O 2.0 If$^\ast$&107 -25/+13&100 -18/+19&107 -25/+13& 94 -12/+10&0.0&2.4&0.1&2.5\\
O 2.0 I& 96 -10/+10& 86 -10/+10& 90 -10/+6& 82 -6/+11&0.8& 2.5&1.1&2.7\\
O 2.5 I& 88 -10/+9& 79 -8/+9& 82 -10/+6& 76 -5/+7&1.0&  2.7&  1.1&  2.8\\
O 3.0 I& 81 -11/+10& 72 -8/+9& 75 -10/+6& 70 -6/+7&1.3&  3.0&  1.3&  3.0\\
O 3.5 I& 74 -11/+12& 66 -8/+8& 68 -9/+6& 64 -6/+8&1.6&  3.1&  1.9&  3.1\\
O 4.0 I& 67 -9/+13& 60 -7/+8& 61 - 7/+7& 59 -6/+8&1.9&  3.2&  2.0&  3.2\\
O 4.5 I& 62 -8/+16& 56 -6/+10& 57 -7/+9& 55 -5/+10&2.3&  3.2&  2.5&  3.3\\
O 5.0 I& 58 -8/+20& 52 -6/+14& 53 -6/+13& 52 -6/+14&2.5&  3.6&  2.8&  3.7\\
O 5.5 I& 55 -9/+22& 49 -7/+16& 50 -7/+15& 49 -6/+16&2.8&  4.0&  2.8&  4.0\\
O 6.0 I& 52 -9/+25& 47 -7/+19& 47 -7/+18& 46 -6/+18&3.1&  4.0&  3.1&  4.1\\
O 6.5 I& 49 -10/+28& 44 -7/+21& 44 -7/+21& 43 -7/+21&3.1&  4.3&  3.1&  4.4\\
O 7.0 I& 46 -9/+30& 41 -7/+23& 42 -7/+22& 41 -6/+23&3.1&  4.7&  3.2&  4.7\\
O 7.5 I& 43 -8/+12& 39 -6/+8& 40 -7/+7& 39 -6/+8&4.2&  5.0&  4.2&  5.1\\
O 8.0 I& 41 -9/+12& 38 -7/+8& 38 -7/+8& 37 -7/+8&4.4&  5.5&  4.4&  5.5\\
O 8.5 I& 39 -9/+13& 36 -7/+9& 36 -7/+9& 36 -7/+9&4.5&  5.9&  4.6&  5.9\\
O 9.0 I& 37 -8/+14& 34 -6/+10& 34 -7/+9& 34 -6/+10&4.6&  6.2&  4.7&  6.3\\
O 9.5 I& 37 -10/+13& 33 -7/+10& 34 -8/+9& 33 -7/+10&4.8&  7.4&  4.8&  7.4\\
\hline
O 2.0 III& 87 -10/+10& 84 -9/+10& 86 -10/+8& 83 -7/+11&0.0&  1.5&  0.1&  1.7\\
O 2.5 III& 78 -10/+12& 76 -9/+11& 77 -11/+9& 75 -8/+12&0.0&  1.8&  0.1&  2.0\\
O 3.0 III& 70 -10/+12& 67 -9/+12& 69 -10/+10& 66 -8/+12&0.1&  2.2&  0.4&  2.2\\
O 3.5 III& 62 -7/+12& 60 -6/+11& 61 -7/+9& 59 -5/+11&0.2&  2.2&  0.6&  2.5\\
O 4.0 III& 57 -6/+9& 54 -5/+9& 55 -5/+7& 54 -4/+9&0.5&  2.5&  1.1&  2.8\\
O 4.5 III& 52 -5/+7& 50 -5/+7& 50 -5/+5& 49 -4/+6&0.8&  2.8&  1.1&  3.0\\
O 5.0 III& 48 -6/+7& 46 -5/+7& 46 -5/+6& 45 -5/+6&1.1&  3.4&  1.6&  3.4\\
O 5.5 III& 44 -5/+7& 42 -4/+7& 42 -4/+6& 41 -4/+7&1.6&  3.7&  1.6&  3.7\\
O 6.0 III& 40 -4/+8& 39 -4/+7& 39 -4/+6& 38 -4/+7&1.9&  3.9&  2.5&  4.0\\
O 6.5 III& 37 -4/+7& 36 -4/+7& 36 -4/+6& 35 -3/+6&2.4&  4.3&  3.0&  4.4\\
O 7.0 III& 34 -4/+5& 33 -4/+5& 33 -4/+4& 33 -3/+4&2.9&  4.4&  3.2&  4.8\\
O 7.5 III& 32 -4/+6& 31 -3/+6& 31 -3/+5& 31 -3/+6&3.4&  4.9&  3.5&  5.3\\
O 8.0 III& 30 -4/+5& 29 -3/+5& 29 -3/+5& 29 -3/+5&3.8&  5.4&  4.7&  5.9\\
O 8.5 III& 28 -4/+5& 27 -3/+5& 27 -4/+5& 27 -3/+5&4.4&  6.1&  5.0&  6.4\\
O 9.0 III& 25 -3/+5& 25 -3/+5& 25 -3/+4& 25 -3/+4&5.4&  7.1&  5.5&  7.1\\
O 9.5 III& 23 -2/+5& 23 -2/+5& 23 -2/+4& 23 -2/+4&5.8&  7.2&  6.2&  7.7\\
\hline
O 2.0 V& 80 -8/+6& 80 -8/+5& 80 -8/+5& 80 -8/+5&0.0&  0.1&  0.1&  0.4\\
O 2.5 V& 72 -9/+6& 72 -9/+6& 72 -9/+5& 72 -9/+6&0.0&  0.4&  0.1&  0.8\\
O 3.0 V& 64 -7/+7& 64 -7/+7& 64 -7/+7& 64 -7/+7&0.0&  0.8&  0.1&  1.1\\
O 3.5 V& 58 -5/+6& 57 -5/+6& 58 -5/+6& 57 -4/+7&0.0&  0.8&  0.1&  1.4\\
O 4.0 V& 53 -5/+4& 53 -5/+4& 53 -5/+4& 52 -5/+4&0.0&  1.2&  0.1&  1.7\\
O 4.5 V& 49 -6/+4& 48 -5/+4& 49 -6/+4& 48 -5/+4&0.0&  2.0&  0.1&  2.0\\
O 5.0 V& 44 -5/+5& 44 -5/+5& 44 -5/+4& 43 -4/+5&0.0&  2.0&  0.1&  2.5\\
O 5.5 V& 40 -5/+4& 40 -5/+4& 40 -5/+4& 39 -4/+4&0.0&  2.1&  0.1&  2.8\\
O 6.0 V& 37 -5/+3& 36 -4/+3& 37 -5/+3& 36 -4/+3&0.0&  3.0&  0.1&  3.2\\
O 6.5 V& 34 -5/+3& 33 -4/+3& 33 -4/+3& 33 -4/+3&0.0&  3.4&  0.1&  3.7\\
O 7.0 V& 31 -5/+3& 30 -4/+3& 31 -5/+3& 30 -4/+3&0.0&  3.6&  0.1&  4.2\\
O 7.5 V& 28 -5/+3& 28 -5/+3& 28 -5/+3& 27 -4/+3&0.0&  4.2&  0.2&  4.8\\
O 8.0 V& 25 -4/+4& 25 -4/+4& 25 -4/+3& 25 -4/+4&0.0&  4.3&  0.1&  5.4\\
O 8.5 V& 22 -4/+5& 22 -4/+5& 22 -4/+4& 22 -4/+4&0.0&  5.8&  0.2&  6.1\\
O 9.0 V& 20 -4/+4& 20 -4/+4& 20 -4/+4& 20 -4/+3&0.0&  6.7&  0.1&  6.8\\
O 9.5 V& 18 -2/+5& 18 -2/+5& 18 -2/+4& 18 -2/+4&0.0&  6.7&  0.1&  7.5\\
\hline
\end{tabular}
\end{table*}

\begin{table*}
\centering
\caption{\label{tab:Oz04} Theoretical masses for O stars from rotating
  SMC (z=0.004) stellar models.}
\begin{tabular}{ccccccccc}
Spectral&$m_\mathrm{ini}$&$m_\mathrm{evol}$&$m_\mathrm{start}$&$m_\mathrm{end}$&
$t_\mathrm{start, min}$& $t_\mathrm{start, max}$& $t_\mathrm{end,
  min}$& $t_\mathrm{end, max}$\\
type&$M_\odot$&$M_\odot$&$M_\odot$&$M_\odot$&Myr&Myr&Myr&Myr\\
\hline
O 2.0 If$^\ast$& 90 -39/+30& 86 -35/+34& 90 -39/+30& 82 -31/+26&0.0&  2.5&  0.1&  2.5\\
O 2.0 I& 97 -20/+23& 86 -17/+21& 87 -17/+20& 86 -16/+20&2.4&  2.6&  2.4&  2.7\\
O 2.5 I& 87 -17/+22& 78 -14/+18& 79 -15/+17& 77 -13/+18&2.6&  2.8&  2.8&  2.9\\
O 3.0 I& 80 -16/+19& 72 -13/+15& 72 -13/+14& 71 -12/+15&2.8&  3.0&  2.9&  3.1\\
O 3.5 I& 75 -16/+18& 67 -12/+14& 67 -13/+13& 67 -12/+14&3.0&  3.1&  3.1&  3.2\\
O 4.0 I& 70 -15/+18& 63 -12/+14& 63 -12/+13& 62 -11/+14&3.1&  3.3&  3.2&  3.4\\
O 4.5 I& 65 -13/+18& 59 -11/+13& 59 -11/+12& 59 -10/+13&3.3&  3.5&  3.3&  3.6\\
O 5.0 I& 61 -13/+19& 55 -10/+14& 55 -10/+14& 55 -10/+14&3.4&  3.8&  3.4&  3.8\\
O 5.5 I& 56 -11/+23& 51 -9/+17& 52 -9/+16& 51 -9/+17&3.4&  4.0&  3.4&  4.1\\
O 6.0 I& 53 -12/+26& 49 -10/+20& 49 -10/+19& 49 -10/+19&3.4&  4.3&  3.4&  4.4\\
O 6.5 I& 51 -12/+27& 47 -10/+21& 47 -10/+20& 47 -10/+20&3.4&  4.6&  3.4&  4.6\\
O 7.0 I& 50 -12/+28& 45 -9/+22& 46 -9/+21& 45 -9/+22&3.5&  4.9&  3.5&  4.9\\
O 7.5 I& 47 -11/+30& 43 -9/+23& 44 -9/+23& 43 -9/+23&3.5&  5.3&  3.5&  5.3\\
O 8.0 I& 46 -12/+14& 42 -10/+11& 42 -10/+11& 42 -10/+11&4.2&  5.5&  4.2&  5.6\\
O 8.5 I& 44 -12/+16& 40 -10/+12& 41 -10/+12& 40 -10/+12&4.3&  5.9&  4.3&  6.0\\
O 9.0 I& 44 -13/+16& 40 -11/+13& 40 -11/+12& 40 -11/+12&4.3&  6.4&  4.3&  6.4\\
O 9.5 I& 41 -12/+18& 38 -10/+14& 38 -10/+14& 38 -10/+14&4.6&  6.8&  4.6&  6.8\\
\hline
O 2.0 III& 71 -11/+17& 67 -10/+15& 67 -10/+14& 66 -9/+15&2.0&  2.5&  2.0&  2.6\\
O 2.5 III& 65 -9/+15& 61 -8/+13& 61 -8/+13& 61 -7/+13&2.2&  2.8&  2.3&  2.8\\
O 3.0 III& 60 -7/+13& 57 -6/+11& 57 -6/+11& 56 -6/+11&2.4&  2.8&  2.7&  3.0\\
O 3.5 III& 56 -7/+10& 53 -6/+9& 53 -6/+9& 52 -6/+9&2.6&  3.0&  2.8&  3.2\\
O 4.0 III& 52 -7/+7& 49 -6/+7& 49 -6/+6& 49 -6/+7&2.9&  3.4&  3.1&  3.4\\
O 4.5 III& 48 -6/+8& 46 -5/+7& 46 -5/+7& 46 -5/+7&3.2&  3.5&  3.3&  3.7\\
O 5.0 III& 45 -6/+8& 43 -5/+8& 43 -5/+7& 42 -5/+7&3.5&  3.7&  3.5&  3.9\\
O 5.5 III& 42 -4/+7& 40 -4/+7& 40 -4/+6& 40 -3/+6&3.8&  4.0&  3.9&  4.2\\
O 6.0 III& 39 -3/+6& 38 -3/+5& 38 -3/+5& 38 -3/+5&4.1&  4.4&  4.1&  4.5\\
O 6.5 III& 37 -3/+5& 36 -3/+5& 36 -3/+4& 35 -3/+5&4.4&  4.8&  4.4&  4.9\\
O 7.0 III& 35 -3/+4& 34 -3/+4& 34 -3/+3& 34 -3/+4&4.7&  5.2&  4.7&  5.2\\
O 7.5 III& 33 -4/+4& 32 -3/+4& 32 -3/+3& 32 -3/+3&5.0&  5.4&  5.1&  5.6\\
O 8.0 III& 32 -4/+4& 30 -3/+4& 30 -3/+4& 30 -3/+4&5.4&  5.9&  5.5&  6.1\\
O 8.5 III& 30 -4/+4& 29 -4/+4& 29 -4/+4& 29 -4/+4&5.7&  6.4&  5.7&  6.5\\
O 9.0 III& 28 -2/+4& 27 -2/+4& 27 -2/+3& 27 -2/+3&6.1&  6.6&  6.1&  6.9\\
O 9.5 III& 27 -7/+4& 26 -6/+4& 26 -6/+4& 26 -6/+4&6.5&  7.4&  6.6&  7.5\\
\hline
O 2.0 V& 57 -10/+13& 56 -9/+12& 56 -9/+11& 56 -9/+12&0.0&  1.9&  0.1&  2.2\\
O 2.5 V& 52 -8/+10& 51 -8/+9& 52 -8/+8& 51 -7/+9&0.0&  2.2&  0.1&  2.4\\
O 3.0 V& 48 -8/+9& 48 -8/+8& 48 -8/+7& 47 -7/+8&0.0&  2.6&  0.1&  2.6\\
O 3.5 V& 45 -7/+9& 44 -6/+9& 44 -6/+8& 44 -6/+9&0.0&  2.7&  0.1&  2.9\\
O 4.0 V& 42 -6/+8& 41 -5/+8& 41 -5/+7& 41 -5/+8&0.0&  2.8&  0.1&  3.1\\
O 4.5 V& 39 -6/+8& 38 -5/+8& 38 -5/+7& 38 -5/+8&0.0&  3.1&  0.2&  3.4\\
O 5.0 V& 36 -5/+7& 35 -5/+7& 36 -5/+6& 35 -4/+6&0.0&  3.5&  0.3&  3.7\\
O 5.5 V& 34 -6/+5& 33 -5/+5& 33 -5/+5& 33 -5/+5&0.0&  4.0&  0.1&  4.1\\
O 6.0 V& 31 -5/+7& 31 -5/+7& 31 -5/+6& 31 -5/+6&0.0&  4.1&  0.3&  4.4\\
O 6.5 V& 29 -5/+7& 29 -5/+7& 29 -5/+6& 28 -5/+6&0.0&  4.4&  0.5&  4.8\\
O 7.0 V& 27 -6/+7& 26 -6/+7& 26 -6/+7& 26 -5/+7&0.0&  5.2&  0.5&  5.2\\
O 7.5 V& 24 -5/+8& 23 -5/+8& 23 -5/+8& 23 -5/+8&0.0&  5.2&  0.1&  5.6\\
O 8.0 V& 21 -4/+8& 21 -4/+8& 21 -4/+8& 21 -3/+8&0.0&  5.6&  1.7&  6.1\\
O 8.5 V& 20 -4/+8& 19 -4/+8& 20 -4/+8& 19 -4/+8&0.0&  6.1&  0.2&  6.7\\
O 9.0 V& 18 -3/+9& 18 -3/+9& 18 -3/+8& 18 -3/+8&0.0&  6.7&  0.9&  7.4\\
O 9.5 V& 16 -3/+7& 16 -3/+7& 16 -3/+7& 16 -3/+7&0.0&  7.4&  0.1&  8.3\\
\hline
\end{tabular}
\end{table*}

The newly arrived spectral-type-mass relation, as arrived in
Sect.~\ref{sec:def}, is now compared with the dynamical
(Table~\ref{tab:dyn}) and literature spectroscopic (Table~\ref{tab:spec})
mass estimates from Sect.~\ref{sec:data}. 

\subsection{Comparison with dynamical masses}
\label{sec:res_dyn}
\begin{figure}
\begin{center}
\includegraphics[width=8cm]{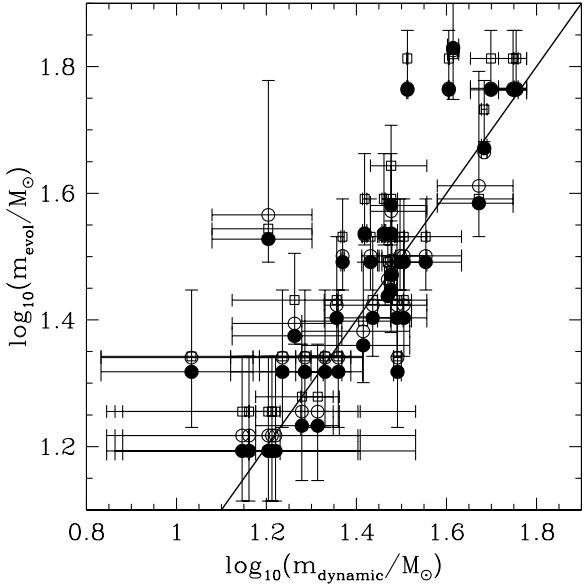}
\vspace*{-2.0cm}
\caption{Comparison of the different mass estimates with dynamically
  determined stellar masses. The {\it solid line} marks a
  one-to-one correspondence between model and dynamical mass. {\it Open
  circles} show the masses for the MSH05 theoretical O star
  \Teff~scale, {\it closed circles} for the MSH05 observational
  temperature scale and {\it open boxes} mark the mean masses derived
  in this work.}
\label{fig:comp_all}
\end{center}
\end{figure}

In Fig.~\ref{fig:comp_all} the results of the MSH05 and the present-day
mass estimates are compared with dynamical mass estimates for massive
stars from the literature as shown in Table~\ref{tab:dyn}. Inspecting
the mass estimates, a very large spread is noticeable.

The results of a linear correlation analysis for all stars in the
sample are shown in Table~\ref{tab:corr}. For each sample, the slope and
offset of a best-fitting linear relation are given together with the correlation
coefficients. The ``MSH1'' column provides the mass estimates
from MSH05 using the theoretical $T_\mathrm{eff}$ scale, whilst ``MSH2''
gives the results from their observational $T_\mathrm{eff}$
scale. ``ini'', ``evol'', ``start'' and ``end'' are the results
arrived at here, with ``ini'' marking the results for initial masses of
the models, ``start'' the evolved mass when a star enters a spectral
type and ``end'' the one when he leaves it. ``Evol'' is the mean mass
computed from $m_\mathrm{start}$ and $m_\mathrm{end}$. 
With an offset very close to 0, a slope of nearly 1, and a
correlation coefficient of $\sim$0.9, the MSH05 mass estimates agree
very well with the dynamical masses of the sample. Furthermore, the
minimal, maximal, and evolutionary present-day masses,
which are calibrated on the theoretical $T_\mathrm{eff}$ scale of
MSH05, agree very well with the observed dynamical masses. The
offsets are quite close to 0 and slopes similarly close to 1,
whilst the correlation coefficients are $\sim$ 0.9 too. The present-day
dynamical masses are therefore quite well reproduced by the
models. Note that the errors for the MSH05 based fits are always 
smaller than the fits with the here-derived values. This is because
the MSH05 values involve no errors, and the fit only contains errors
in the dynamical masses. The values derived here values also have
their own error estimate.

In Figs.~\ref{fig:comp_I} to \ref{fig:comp_V} the dynamical masses
from Table~\ref{tab:dyn} (open circles) are shown together with the
present-day mass range ($m_\mathrm{start}$ to $m_\mathrm{end}$) for
the models (shaded region), for the supergiants
(Fig.~\ref{fig:comp_I}), giants (Fig.~\ref{fig:comp_III}) and dwarfs
(Fig.~\ref{fig:comp_V}). Evidently, most of the
dynamical measurements agree well with the models within the error
bars. For supergiants and giants the mass ranges for different
metallicities are nearly indistinguishable. Only the non-rotating
evolutionary models stand out in the case of the supergiants. Because the
dependence of mass loss on rotation is not understood very well, 
this discrepancy is likely very dependent on the assumptions in the
stellar evolutionary code. 
For dwarf stars only the SMC metallicity mass ranges differ visibly
from the solar and LMC estimates. Somewhat surprisingly, the
early-type SMC dwarfs seem to have {\it lower} masses than their MW
and LMC cousins, according to the models and definitions used
here. This is almost certainly because of the earlier mentioned 
perhaps unexpected fact that the \citet{HLH06} SMC O stars result in a
lower \Teff~scale for O stars than for LMC objects.

Interestingly, all but two of the (dwarf) stars located in the LMC 
({\it en-circled circles} in Fig.~\ref{fig:comp_V})
lie below the here derived solar and LMC metallicity
evolutionary mass ranges, independent of rotation and
metallicity. Only the allowed mass range for SMC metallicity
covers these stars.

This might be because of binary stellar evolution because it is more difficult
to access if the stars are detached or not in the LMC. The lower
metallicity of the LMC might also not be accounted for correctly
either in the evolutionary models or in the atmosphere models. The
large distance to these stars compared to the rest of the sample might
also influence the observed values. However, removing these nine stars
from the already small sample of only 30 stars would strongly reduce
its significance.

In one case (V1182 Aql A) the initial mass is slightly
{\it above} the dynamical mass for its spectral type, even when
considering the uncertainties in the observational and 
model mass determination. This might be caused by a somewhat optimistic
observational error ($\pm$ 0.6 $M_\odot$) or can be because of binary
stellar evolution effects such as mass transfer, or excess irradiation
of one stellar hemisphere in tidally locked configurations.

\begin{table*}
\centering
\caption{\label{tab:corr} Correlation function values.}
\begin{tabular}{lcrrccc}
Value&$m_\mathrm{MSH1}$&$m_\mathrm{MSH2}$&$m_\mathrm{ini}$&$m_\mathrm{evol}$&$m_\mathrm{start}$&$m_\mathrm{end}$\\
\hline
offset      &0.017 $\pm$ 0.300&-0.051 $\pm$ 0.311& 0.016 $\pm$ 0.308&0.052 $\pm$ 0.296&0.027 $\pm$ 0.295&0.053 $\pm$ 0.269\\
slope       &1.010 $\pm$ 0.207& 1.048 $\pm$ 0.214& 1.043 $\pm$ 0.216&1.001 $\pm$ 0.206&1.029 $\pm$ 0.209&1.008 $\pm$ 0.208\\
corr. coeff.&0.887 $\pm$ 0.118& 0.879 $\pm$ 0.115& 0.871 $\pm$ 0.184&0.858 $\pm$ 0.174&0.854 $\pm$ 0.170&0.855 $\pm$ 0.173\\
\hline
\end{tabular}
\end{table*}

\subsection{Comparison with spectroscopic masses}
\label{sec:res_spec}

\begin{figure}
\begin{center}
\includegraphics[width=8cm]{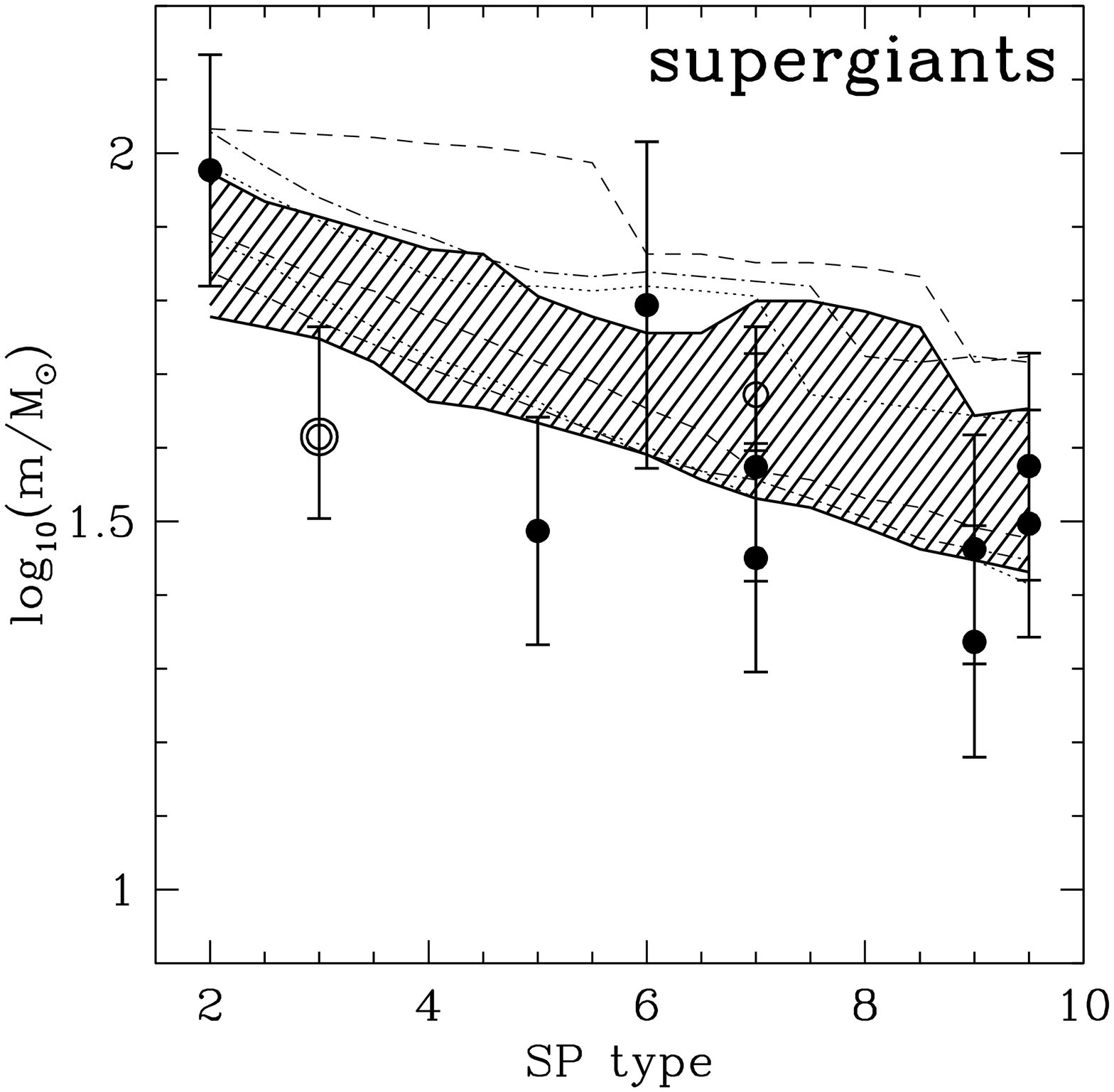}
\vspace*{-2.0cm}
\caption{Comparison of the dynamical ({\it open circles}) and
  spectroscopic ({\it filled circles}) mass measurements for
  luminosity class I objects (supergiants) with the
  evolutionary masses. The {\it shaded region} between the {\it solid
    lines} shows the full range ($m_\mathrm{start}$ to
  $m_\mathrm{end}$) of evolutionary masses for the rotating solar
  metallicity models. The {\it dashed lines} mark the upper and lower
  mass ranges for the non-rotating solar metallicity models, while the {\it
    dotted lines} bracket the possible masses for the rotating LMC
  metallicity models and the {\it dashed-dotted lines} enclose the SMC
  metallicity models. {\it Double encircled} objects are located in the LMC.} 
\label{fig:comp_I}
\end{center}
\end{figure}

\begin{figure}
\begin{center}
\includegraphics[width=8cm]{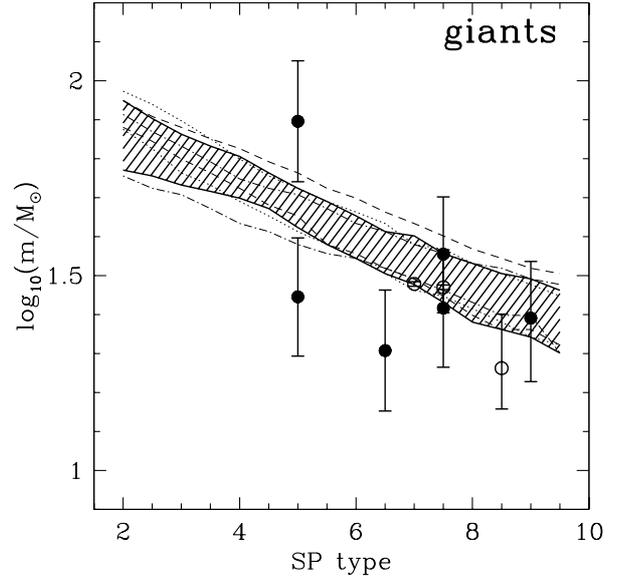}
\vspace*{-2.0cm}
\caption{Like Fig.~\ref{fig:comp_I} but for luminosity class III stars
(giants).}
\label{fig:comp_III}
\end{center}
\end{figure}

\begin{figure}
\begin{center}
\includegraphics[width=8cm]{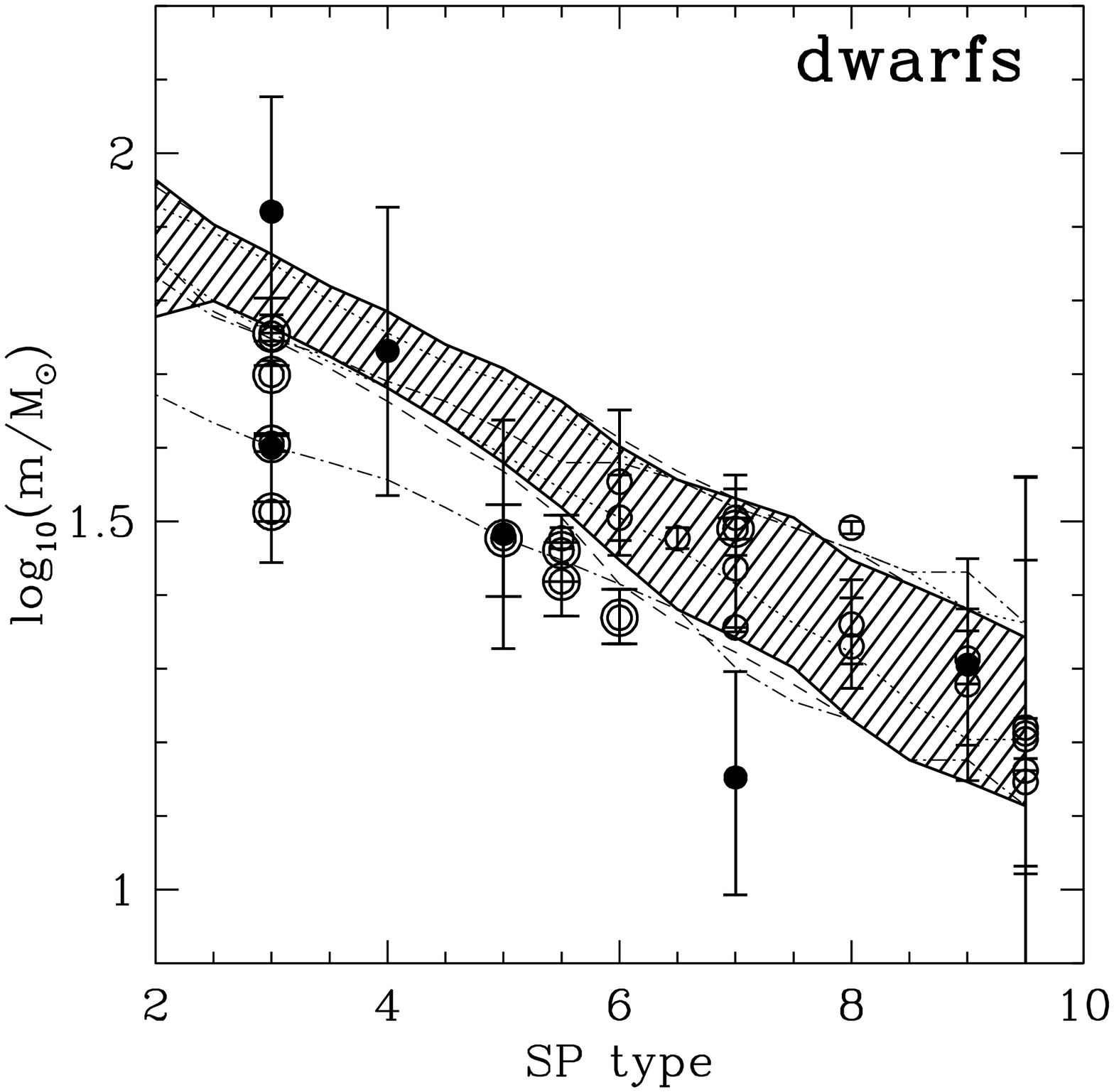}
\vspace*{-2.0cm}
\caption{Like Fig.~\ref{fig:comp_I} but for luminosity class V stars
(dwarfs).}
\label{fig:comp_V}
\end{center}
\end{figure}

Spectroscopic mass determinations are the only other empirical 
method to derive stellar masses for single stars and non-eclipsing
binaries. The hot and usually rapidly-rotating O stars have very broad
spectral features, which often result in large error bars from the
line-fitting techniques used to fit model spectra to observations. The
literature spectroscopic mass values from \citet{RPH04} are shown in
Table~\ref{tab:spec} and are also plotted in Figs.~\ref{fig:comp_I}
to \ref{fig:comp_V} as {\it filled circles}. The spread seems to be
somewhat larger than the spread of the dynamical mass
estimates and it should be noted that several (3 out of 21) of the
spectroscopic measurements are so far outside the predictions that the
error bars do not overlap. Two of these three stars are giants
(HD190864 and HD193682) and one is a dwarf (HD217086). Reassuringly,
all supergiants (Fig.~\ref{fig:comp_I}) overlap at least with their
error bars with the model predictions, and apart from the mentioned
exceptions all giants and dwarfs, too.

The large spread and the three outlying stars might be because of
the generally much larger errors of the spectroscopic masses
determinations compared to dynamical ones. \citet{HLH06} find in
their study of SMC stars that the mass discrepancy is reduced but not
eliminated when using NLTE line-blanketed model atmospheres. They
suggest that the remaining difference is actually a signature of fast
rotation that would lower the apparent surface gravity. As also
indicated in Table~\ref{tab:spec}, some of the stars are known
binaries and two runaway stars are also included in the sample, though
theses stars agree reasonably well with the models. For such objects
binary stellar evolution could have strongly influenced the
present-day mass, and/or spectrum of the star. Nonetheless, the
evolutionary model based spectral-type--mass-relation derived here
agrees with dynamical as well as spectroscopic mass estimates,
suggesting the systematic mass-discrepancy problem might be solved.

\section{Summary and conclusions}
\label{sec:summ}
With the spectral type definition of MSH05, a relation 
between O star spectral type and its mass and age was derived for
solar metallicity stars. This was achieved by taking stellar
evolution models \citep{MM03} and comparing their output luminosities
and \Teff~with the MSH05 spectral type definitions. With the
\citet{MKE07} study of O stars in the LMC, the \citet{HLH06} data
for SMC O stars and evolutionary tracks by \citet{MM05} and
\citet{BNG09}, similar spectral-type--mass-relations were derived for
LMC and SMC metallicity.
The resulting mass versus spectral-type relation agrees well
with dynamical as well as spectroscopic mass measurements from the
literature for MW and LMC main-sequence stars. For SMC stars, giants
and supergiants too few or no dynamical mass estimates are available
for an in-depth comparison. 

Tables~\ref{tab:Orot} and \ref{tab:Onorot} provide easy access to the
mass estimates for a given spectral type based on rotating
(Table~\ref{tab:Orot}) and non-rotating (Table~\ref{tab:Onorot}) solar
metallicity O star models. Tables~\ref{tab:Oz08} and
\ref{tab:Oz04} provide the same for LMC and SMC metallicity.

Furthermore, the evolution of individual stellar models
\citep{MM03,MM05,BNG09} through the different spectral types are given
in appendix~\ref{app:stevol}. For the z = 0.02 and 0.04 the MSH05 solar
metallicity definitions of spectral types are used, while for $z$ =
0.008 the LMC ones and for $z$ = 0.004 the SMC ones.

The new calibration of O star spectral types presented here
with stellar evolution models provides a valuable new tool to derive
O star masses, including initial and present-day masses, and
includes an estimate of the errors. 
Furthermore, the minimal and maximal start and end ages for a given
spectral class provide relevant information for statistical studies of
roughly solar-metallicity young stellar populations.
The relation derived here between spectral type and evolutionary mass
agrees very well with dynamical as well as spectroscopic mass
estimates for O stars from the literature and is therefore quite
robust. No systematic discrepancy between dynamical, model and
spectroscopic mass estimates could be found.

Because there are still considerable error margins in the new calibration,
more observational and theoretical effort is necessary in order to
improve on these. Larger samples of O stars with homogeneously derived
parameters and larger sets of evolutionary models with a broader range
of initial conditions ($v_\mathrm{rot}$, metallicity, magnetic fields
and, especially initial mass) would help to improve the
calibration. Interestingly, seven out of nine stars located in the LMC ({\it
en-circled circles} in Figs.~\ref{fig:comp_I} and \ref{fig:comp_V})
are below or at the lower end of the predicted evolutionary mass
range. Even when considering an LMC metallicity luminosity-\Teff~grid
and evolutionary models ({\it dashed lines} in Figs.~\ref{fig:comp_I} to
\ref{fig:comp_V}). These systematically lower dynamic masses could
have several reasons. The influence of metallicity on the atmosphere
and stellar models might be underestimated or these stars could be
undetected contact systems instead of detached systems. Or stars
earlier than O6 in the LMC are systematically misclassified and 
should all be shifted by one spectral subtype towards later types. 

An additional startling fact are the low \Teff~of the spectral type 
definitions for SMC metallicities and the resulting lower masses for
early SMC O stars. This is probably because of the lack of early O
stars in the \citet{HLH06} SMC study used here to calibrate the SMC
spectral types. Yet these lower masses fit the eclipsing LMC (!) early
O stars much better than the hot LMC spectral type definitions.

The other major source for uncertainties is massive binary 
evolution. It is a major obstacle, especially for giant and supergiant
systems with dynamical mass estimates. These are generally
short-period systems and the increasing radii of giants and
supergiants during their evolution make mass transfer highly likely.
Whilst the modelling of the relevant processes has significantly
improved in recent years \citep{EIT08,LCY08,DCL09,SZL09,VDM09}, the huge
parameter space of hitherto unknown initial conditions (mass ratio,
eccentricity, period, orbital inclination) and the possibility of
reaching the same final state from different initial ones makes a
correction for binary stellar evolutionary effects most challenging.

Given the internal consistency of the three 
model-dependent mass determinations ($M_{\rm evol}$, $M_{\rm spec}$
and $M_{\rm wind}$) it would be tempting to conclude that the basic properties 
of main-sequence O-type stars are now well understood. However, we
point out that all that wind masses, the MSH05 calibrations, and
the rotating Geneva stellar tracks all employ the same underlying mass-loss 
prescriptions of \citet{VDL00}, and these have recently been suggested 
to be too high \citep[e.g.][]{FMP06} as a result of wind clumping. 
If calls for a fundamental downwards revision for O-star mass-loss rates were 
proven to be correct, this would undoubtedly result in the creation 
of new mass discrepancies. It may therefore be considered 
highly significant that current model masses seem to be backed up by 
model-independent dynamical masses -- boosting confidence in our  
basic knowledge of the physical parameters, such as its mass-loss rate, and 
the modelling of the atmospheres and evolution of O-type stars. 

Ironically it appears that the O-stars are currently better 
understood than the adjacent spectral 
type B-type stars, for which \citet[][see their Fig.~10; with data
  from \citet{TDH07} and \citet{HLD08}]{CLB09} showed 
a highly significant mass discrepancy, with evolutionary masses 
up to a factor three larger than spectroscopic ones. 

\begin{acknowledgements}
CW is happy to thank Jan Pflamm-Altenburg, Jim Dale, Nick Moeckel and
Ian Bonnell for helpful discussions. This work made use of the Simbad
web based database. This work was financially supported by the
CONSTELLATION European Commission Marie Curie Research Training
Network (MRTN-CT-2006-035890).
\end{acknowledgements}

\begin{appendix}
\section{Stellar evolution}
\label{app:evol}

The evolution of stars used in this work is based on the stellar
evolution models for solar and non-solar metallicity, rotating with
300 km/s and non-rotating by \citet{MM03} and
\citet{MM05} and each two models with LMC and SMC metallicity (15 and 20
$M_\odot$) by \citet{BNG09}. Model tracks are only provided for stars
with 9, 12, 15, 20, 25, 40, 60, 85 and 120 $M_\odot$ for solar
metallicity and even fewer for non-solar metallicity. As the lifetime
of the stars and their respective evolutionary stages are dependent on
the mass of the star, it is not possible to linearly interpolate
between the track of a 40 and a 60 $M_\odot$ star in order to get, for
example, a 50 $M_\odot$ star. Therefore, a special interpolation
routine is employed here. The model tracks immediately above and below
the target mass are normalized to their individual lifetimes (the
point when the star becomes a neutron star or a black hole). Then the
two normalized tracks are interpolated linearly to the target mass.
The resulting track is then multiplied with the lifetime for the
targeted star. This lifetime is linearly interpolated from the
lifetime of the two input models.

\begin{figure*}
\begin{center}
\includegraphics[width=16cm]{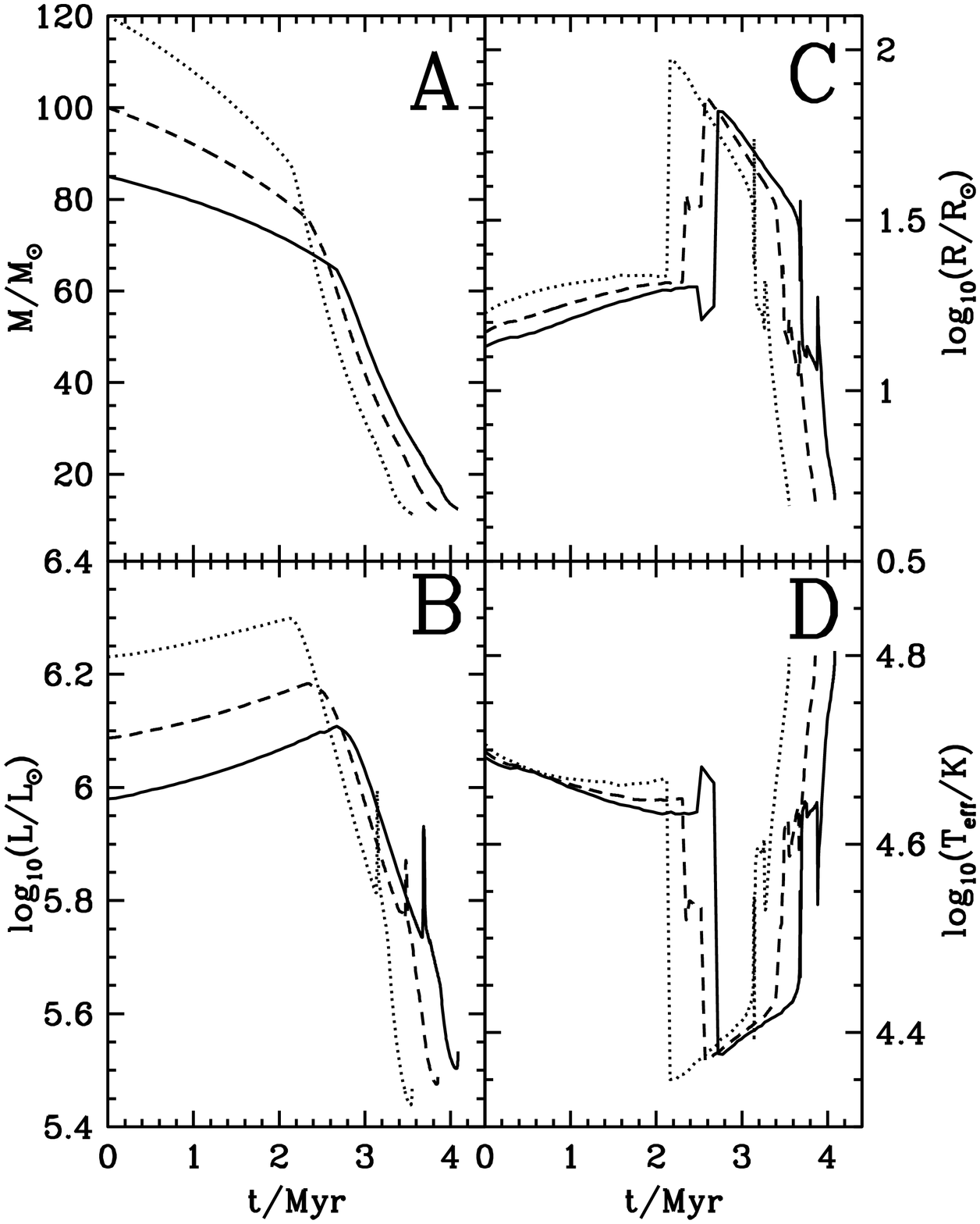}
%\vspace*{-1.0cm}
\caption{{\it Panel A}: Mass evolution over 4 Myr for a 85
  $M_{\odot}$ ({\it solid line}) and a 120 $M_{\odot}$ star ({\it
  dotted line}) from literature data \citep{MM03}. The {\it dashed
    line} shows a 100 $M_{\odot}$ 
  star interpolated from the two models. {\it Panel B}: Luminosity
  evolution over 4 Myr for the same three stars as in the {\it panel
    A}. {\it Panel C}: Radius evolution over 4 Myr for the same three
  stars as in {\it panel A}. {\it Panel D}: Effective temperature
  evolution over 4 Myr for the same three stars as in {\it panel A}.
}
\label{fig:model100ml}
\end{center}
\end{figure*}

Figure~\ref{fig:model100ml} shows the stellar evolution of a 85
$M_\odot$ and a 120 $M_\odot$ star with time from \citet{MM03} for
several stellar parameters (luminosity, radius, mass,
\Teff) together with an interpolated track of a 100 $M_\odot$ star.

\section{Spectral-type--stellar evolution tables}
\label{app:stevol}

The following Table~\ref{tab:MM03z20} shows the different spectral
types that stellar evolution models reach during their lifetime. It uses
solar metallicity \citep{MM03} and non-solar metallicities
\citep{MM05} and rotating ($v_\mathrm{rot, initial}$ = 300 km/s)
and non-rotating tracks between 20 and 120 $M_\odot$.
In this table are also shown evolutionary phases that go beyond
the O spectral type used in the current work. 
Below we describe how these additional phases are
classified. As the surface abundances for several species (H,
He, C, N, O, Ne and Al) are also given in the models, it is possible to
assign the beginning of the hydrogen-rich Wolf-Rayet phase (WNL) as soon
the surface hydrogen abundance is below 60\% \citep{HGL06} and the
helium-rich Wolf-Rayet phase (WNE) is given when the surface
abundance of hydrogen is below $10^{-4}$. Later on, stars are
designated as carbon-rich Wolf-Rayet stars (WC) when helium starts to
be depleted on the surface and the carbon abundance rises above $10^{-4}$.
Exceptions from this scheme are made when the stars enter the
Luminous Blue Variable (LBV), Yellow Hyper-Giant (YHG), Yellow Giant
(YG),  Blue Supergiant (BSG) or Red Supergiant (RSG) phases. The hot
end of the LBVs is defined by $\log_{10}(L_\mathrm{LBV~hot}) = 2.2056 \cdot
\log_{10}(T_\mathrm{eff}) - 3.7737$ and on the cool edge by
$T_\mathrm{LBV~cool}$ = 7500 K with a lower limit of
$\log_{10}(L/L_\odot)$ = 5.3 \citep{SVD04}. YHGs lie between 4500 to
7500 K and $\log_{10}(L/L_\odot)$ $\ge$ 5.3 \citep{SVD04}. Models
that evolve between 4500 to 7500 K, but have $\log_{10}(L/L_\odot)$
$<$ 5.3, are named YGs. The BSGs are stars that are too cold for the
O9.5 III or the O9.5 I types, but that are still below the LBV limit
and hotter than YHGs. And finally, RSGs are stars colder than 4500 K
and $\log_{10}(L) \ge$ 3.55 $\log_{10}(L/L_\odot)$ \citep{Le09b}.

The evolutionary sequence for solar metallicity roughly agrees with
the currently used observational sequence by e.g. \citet{Cr07}:
\begin{itemize}
  \item $M_\mathrm{initial} >$ 75 $M_\odot$: O $\rightarrow$ WNL
    $\rightarrow$ LBV $\rightarrow$ WNE $\rightarrow$ WC
    $\rightarrow$ SNIc,
  \item $M_\mathrm{initial} =$ 40 - 75  $M_\odot$: O $\rightarrow$ LBV
    $\rightarrow$ WNE $\rightarrow$ WC $\rightarrow$ SNIc,
  \item $M_\mathrm{initial} =$ 25 - 40  $M_\odot$: O $\rightarrow$
    LBV/RSG $\rightarrow$ WNE $\rightarrow$ SNIb.
\end{itemize}
Where SNIb and SNIc are supernovae type Ib and Ic, although it has recently
been suggested that LBVs could explode early \citep{KV06,GL09}, 
which would significantly alter the later evolutionary phases of these
types of schemes.

\end{appendix}

\bibliographystyle{aa}
\bibliography{mybiblio}

\clearpage \onecolumn
\setcounter{table}{8}
\begin{longtable}{ccccc|ccccc}
%{\small
\caption{\label{tab:MM03z20} Spectral type evolution of stellar models.}\\
\hline\hline
Age&mass&Luminosity&\Teff&Sp.~Type&Age&mass&Luminosity&\Teff&Sp.~Type\\
Myr&$M_\odot$&$\log_{10}(L/L_\odot)$&$\log_{10}(\mathrm{K})$&&Myr&$M_\odot$&$\log_{10}(L/L_\odot)$&$\log_{10}(\mathrm{K})$&\\
\hline
\endfirsthead
\caption{continued.}\\
\hline\hline
Age&mass&Luminosity&\Teff&Sp.~Type&Age&mass&Luminosity&\Teff&Sp.~Type\\
Myr&$M_\odot$&$\log_{10}(L/L_\odot)$&$\log_{10}(\mathrm{K})$&&Myr&$M_\odot$&$\log_{10}(L/L_\odot)$&$\log_{10}(\mathrm{K})$&\\
\hline
\endhead
\hline
\endfoot
\multicolumn{10}{c}{solar metallicity ($z$ = 0.02)}\\
\hline
\multicolumn{5}{c|}{$m_\mathrm{ini}$ = 120 $M_\odot$, $v_\mathrm{rot}$
= 300 km/s, $z$ = 0.02}&\multicolumn{5}{c}{$m_\mathrm{ini}$ = 120
  $M_\odot$, $v_\mathrm{rot}$ = 0 km/s, $z$ = 0.02}\\
 0.0000000 & 120.0000 &  6.2310 &  4.7050 & O 2.0 If$^\ast$&0.0000000 & 120.0000 &  6.2340 &  4.7120 & O 2.0 If$^\ast$\\
 1.3778912 & 102.0845 &  6.2700 &  4.6660 & WNL&1.1227475 & 107.0192 &  6.2530 &  4.6420 & O 2.0 I\\
 2.1617455 &  86.5949 &  6.2980 &  4.3500 & LBV&1.2093965 & 105.8202 &  6.2550 &  4.6320 & O 2.5 I\\
 2.8960620 &  36.2371 &  5.9300 &  4.4030 & WNL&1.2941639 & 104.6224 &  6.2580 &  4.6220 & O 3.0 I\\
 3.1424342 &  26.0184 &  5.9820 &  4.4230 & LBV&1.4177116 & 102.8426 &  6.2620 &  4.6050 & O 3.5 I\\
 3.1458282 &  25.8019 &  5.9880 &  4.4350 & WNL&1.4978035 & 101.6792 &  6.2650 &  4.5930 & O 4.0 I\\
 3.1906745 &  23.4728 &  5.8080 &  4.5920 & WNE&1.6138339 &  99.9870 &  6.2700 &  4.5790 & O 4.5 I\\
 3.2641200 &  20.3641 &  5.7360 &  4.5750 &  WC&1.7257569 &  98.3582 &  6.2750 &  4.5630 & O 5.0 I\\
 3.5503435 &  11.2971 &  5.4710 &  4.7980 &  WC&1.7623851 &  97.8324 &  6.2770 &  4.5560 & LBV\\
&&&&&2.7496870 &  41.2601 &  6.1690 &  4.5080 & WNL\\
&&&&&2.7713165 &  38.9317 &  6.1310 &  4.5120 & WNE\\
&&&&&2.7989238 &  36.2017 &  6.0930 &  4.5090 &  WC\\
&&&&&3.0980530 &  16.2928 &  5.7460 &  4.8130 &  WC\\
\hline
\multicolumn{5}{c|}{$m_\mathrm{ini}$ = 85 $M_\odot$, $v_\mathrm{rot}$
= 300 km/s, $z$ = 0.02}&\multicolumn{5}{c}{$m_\mathrm{ini}$ = 85
  $M_\odot$, $v_\mathrm{rot}$ = 0 km/s, $z$ = 0.02}\\
 0.0000000 &  85.0000 &  5.9800 &  4.6910 & O 2.0 V&0.0000000 & 85.0000 &  5.9840 &  4.6990 & O 2.0 If$^\ast$\\
 1.0302848 &  79.4176 &  6.0160 &  4.6590 & O 2.0 III&0.1533940 &  84.3027 &  5.9870 &  4.6900 & O 2.0 V\\
 1.4552295 &  76.4990 &  6.0360 &  4.6450 & O 2.5 I&0.9607395 &  80.0894 &  6.0100 &  4.6600 & O 2.0 III\\
 2.0694338 &  71.3398 &  6.0710 &  4.6320 & WNL&1.2053038 &  78.6074 &  6.0190 &  4.6470 & O 2.5 III\\
 2.7229040 &  62.8013 &  6.1040 &  4.3780 & LBV&1.2565218 &  78.2833 &  6.0200 &  4.6440 & O 2.5 I\\
 3.1913865 &  40.6504 &  5.9420 &  4.4060 & WNL&1.4553924 &  76.9794 &  6.0280 &  4.6310 & O 3.0 I\\
 3.7471715 &  21.0334 &  5.7310 &  4.6420 & WNE&1.5975672 &  76.0046 &  6.0340 &  4.6190 & O 3.5 I\\
 3.8772150 &  17.3227 &  5.6510 &  4.6420 &  WC&1.6891948 &  75.3598 &  6.0380 &  4.6090 & O 4.0 I\\
 4.0856488 &  12.3616 &  5.5330 &  4.8050 &  WC&1.8224420 &  74.4070 &  6.0440 &  4.5940 & O 4.5 I\\
&&&&&1.9502418 &  73.4820 &  6.0510 &  4.5800 & O 5.0 I\\
&&&&&2.0732452 &  72.5979 &  6.0570 &  4.5600 & O 5.5 I\\
&&&&&2.1521198 &  72.0525 &  6.0620 &  4.5410 & O 6.0 I\\
&&&&&2.2289090 &  71.5513 &  6.0670 &  4.5260 & O 6.5 I\\
&&&&&2.3037808 &  71.0943 &  6.0730 &  4.5050 & O 7.0 I\\
&&&&&2.4116148 &  70.5319 &  6.0810 &  4.4690 & BSG\\
&&&&&2.4464302 &  70.3800 &  6.0840 &  4.4580 & LBV\\
&&&&&3.1313858 &  39.1867 &  6.1750 &  4.5230 & WNL\\
&&&&&3.1693195 &  36.0928 &  6.0530 &  4.5900 & WNE\\
&&&&&3.1985408 &  33.6245 &  6.0170 &  4.5210 &  WC\\
&&&&&3.4635015 &  17.2654 &  5.7590 &  4.8010 &  WC\\
\hline
\multicolumn{5}{c|}{$m_\mathrm{ini}$ = 60 $M_\odot$, $v_\mathrm{rot}$
= 300 km/s, $z$ = 0.02}&\multicolumn{5}{c}{$m_\mathrm{ini}$ = 60
  $M_\odot$, $v_\mathrm{rot}$ = 0 km/s, $z$ = 0.02}\\
 0.0000000 &  60.0000 &  5.7020 &  4.6680 & O 3.0 V&0.0000000 &  60.0000 &  5.7080 &  4.6750 & O 3.0 V\\
 0.1958980 &  59.6226 &  5.7080 &  4.6620 & O 3.5 V&0.5715711 &  58.8823 &  5.7270 &  4.6590 & O 3.5 V\\
 1.3372419 &  56.9559 &  5.7600 &  4.6350 & O 4.0 V&1.3906442 &  56.9734 &  5.7600 &  4.6350 & O 4.0 V\\
 1.6336100 &  56.0998 &  5.7760 &  4.6280 & O 4.0 III&1.5717474 &  56.4936 &  5.7680 &  4.6280 & O 4.0 III\\
 2.4737502 &  53.1420 &  5.8290 &  4.6100 & O 4.5 III&1.8594586 &  55.6850 &  5.7810 &  4.6140 & O 4.5 III\\
 2.7917235 &  51.7499 &  5.8530 &  4.6050 & O 4.5 I&2.1249805 &  54.8898 &  5.7950 &  4.5980 & O 5.0 III\\
 2.9710480 &  50.8750 &  5.8660 &  4.6030 & WNL&2.2265692 &  54.5749 &  5.8000 &  4.5900 & O 5.0 I\\
 3.9396822 &  44.1082 &  5.9510 &  4.3440 & LBV&2.3720850 &  54.1169 &  5.8090 &  4.5770 & O 5.5 I\\
 4.2857995 &  32.3067 &  5.9200 &  4.3970 & WNL&2.5574298 &  53.5309 &  5.8200 &  4.5570 & O 6.0 I\\
 4.3073855 &  31.2900 &  6.0210 &  4.4370 & LBV&2.6887792 &  53.1230 &  5.8290 &  4.5400 & O 6.5 I\\
 4.3082900 &  31.2285 &  6.0620 &  4.4700 & WNL&2.8151875 &  52.7477 &  5.8380 &  4.5200 & O 7.0 I\\
 4.3089005 &  31.1840 &  6.0690 &  4.4620 & LBV&2.9347828 &  52.4192 &  5.8470 &  4.4980 & O 7.5 I\\
 4.3197685 &  30.3332 &  6.0900 &  4.4730 & WNL&3.0123325 &  52.2268 &  5.8530 &  4.4800 & O 8.0 I\\
 4.3670145 &  27.5482 &  5.9130 &  4.6090 & WNE&3.0857660 &  52.0650 &  5.8590 &  4.4610 & O 8.5 I\\
 4.4334785 &  24.1890 &  5.8570 &  4.5920 &  WC&3.1218755 &  51.9937 &  5.8620 &  4.4500 & O 9.0 I\\
 4.6809210 &  14.6686 &  5.6460 &  4.7990 &  WC&3.1912615 &  51.8739 &  5.8680 &  4.4280 & O 9.5 I\\
&&&&& 3.2612565 &  51.2141 &  5.8720 &  4.3950 & BSG\\
&&&&& 3.2947412 &  50.6178 &  5.8730 &  4.3720 & LBV\\
&&&&& 3.6153805 &  33.7033 &  5.8670 &  4.3750 & BSG\\
&&&&& 3.6224198 &  33.5101 &  5.8940 &  4.4460 & O 9.0 I\\
&&&&& 3.6229290 &  33.5080 &  5.9030 &  4.4550 & O 8.5 I\\
&&&&& 3.6233745 &  33.5060 &  5.9080 &  4.4530 & O 9.0 I\\
&&&&& 3.6240875 &  33.5032 &  5.9100 &  4.4300 & O 9.5 I\\
&&&&& 3.6245332 &  33.4851 &  5.9090 &  4.3970 & BSG\\
&&&&& 3.6246225 &  33.4817 &  5.9100 &  4.3840 & LBV\\
&&&&& 3.6258830 &  33.1985 &  5.9280 &  3.8730 & YHG\\
&&&&& 3.6274002 &  32.9562 &  5.9930 &  3.9290 & LBV\\
&&&&& 3.6279472 &  32.7957 &  5.9760 &  3.8650 & YHG\\
&&&&& 3.6408085 &  28.8736 &  6.0340 &  3.8770 & LBV\\
&&&&& 3.6727700 &  26.6576 &  5.9840 &  4.4310 & WNL\\
&&&&& 3.7211365 &  24.5152 &  5.8540 &  4.6000 & WNE\\
&&&&& 3.8060685 &  21.0768 &  5.7380 &  4.6670 &  WC\\
&&&&& 4.0147065 &  14.6174 &  5.6480 &  4.7950 &  WC\\
\hline
%}
\multicolumn{5}{c|}{$m_\mathrm{ini}$ = 40 $M_\odot$, $v_\mathrm{rot}$
= 300 km/s, $z$ = 0.02}&\multicolumn{5}{c}{$m_\mathrm{ini}$ = 40
  $M_\odot$, $v_\mathrm{rot}$ = 0 km/s, $z$ = 0.02}\\
 0.0000000 &  40.0000 &  5.3410 &  4.6290 & O 5.0 V&0.0000000 &  40.0000 &  5.3500 &  4.6380 & O 5.0 V\\
 0.6033284 &  39.6638 &  5.3650 &  4.6190 & O 5.5 V&1.2306326 &  39.2770 &  5.3980 &  4.6160 & O 5.5 V\\
 2.7180158 &  37.9967 &  5.4770 &  4.5830 & O 6.0 V&2.5606410 &  38.2461 &  5.4630 &  4.5830 & O 6.0 V\\
 2.8079328 &  37.9023 &  5.4830 &  4.5800 & O 6.0 III&2.6943652 &  38.1257 &  5.4710 &  4.5780 & O 6.0 III\\
 3.5662060 &  37.0044 &  5.5360 &  4.5560 & O 6.5 III&3.0685855 &  37.7742 &  5.4920 &  4.5600 & O 6.5 III\\
 3.9535328 &  36.4431 &  5.5660 &  4.5400 & O 7.0 III&3.3507872 &  37.4991 &  5.5100 &  4.5430 & O 7.0 III\\
 4.1738048 &  36.0534 &  5.5850 &  4.5310 & O 7.0 I&3.6131688 &  37.2431 &  5.5280 &  4.5220 & O 7.5 III\\
 4.4472590 &  35.5514 &  5.6090 &  4.5170 & O 7.5 I&3.6630180 &  37.1954 &  5.5310 &  4.5170 & O 7.5 I\\
 4.7574220 &  34.9711 &  5.6390 &  4.4950 & O 8.0 I&3.8064682 &  37.0616 &  5.5420 &  4.5020 & O 8.0 I\\
 4.9266050 &  34.6601 &  5.6570 &  4.4780 & O 8.5 I&3.9426182 &  36.9415 &  5.5530 &  4.4850 & O 8.5 I\\
 5.0309235 &  34.4764 &  5.6690 &  4.4660 & O 9.0 I&4.0279992 &  36.8714 &  5.5600 &  4.4720 & O 9.0 I\\
 5.1301465 &  34.3095 &  5.6800 &  4.4530 & WNL&4.1534765 &  36.7774 &  5.5710 &  4.4530 & O 9.5 I\\
 5.5382560 &  32.8282 &  5.7780 &  4.3160 & LBV&4.3112995 &  36.6785 &  5.5860 &  4.4220 & BSG\\
 5.5404720 &  32.5145 &  5.8030 &  3.8580 & YHG&4.5648590 &  35.3579 &  5.6560 &  4.2690 & LBV\\
 5.6038870 &  23.6730 &  5.8910 &  4.2830 & LBV&4.5651770 &  35.3557 &  5.6590 &  4.2420 & BSG\\
 5.6318410 &  22.4679 &  5.8770 &  4.3820 & WNL&4.5681685 &  35.1705 &  5.6620 &  3.8600 & YHG\\
 5.7271790 &  19.2964 &  5.7050 &  4.6260 & WNE&4.9779120 &  15.8704 &  5.6140 &  3.8970 & LBV\\
 5.7341225 &  19.0507 &  5.7060 &  4.6100 &  WC&4.9839990 &  15.4650 &  5.6120 &  4.5060 & WNL\\
 5.9676575 &  12.7371 &  5.5540 &  4.8000 &  WC&5.0102080 &  15.0283 &  5.6100 &  4.6520 & WNE\\
 &&&&&5.0534350 &  14.0897 &  5.6210 &  4.6460 & WNE\\
 \hline
\multicolumn{5}{c|}{$m_\mathrm{ini}$ = 25 $M_\odot$, $v_\mathrm{rot}$
= 300 km/s, $z$ = 0.02}&\multicolumn{5}{c}{$m_\mathrm{ini}$ = 25
  $M_\odot$, $v_\mathrm{rot}$ = 0 km/s, $z$ = 0.02}\\
 0.0000000 &  25.0000 &  4.8620 &  4.5660 & O 7.0 V&0.0000000 &  25.0000 &  4.8730 &  4.5760 & O 6.5 V\\
 2.5403295 &  24.7687 &  4.9570 &  4.5460 & O 7.5 V&0.7044491 &  24.9440 &  4.8940 &  4.5670 & O 7.0 V\\
 4.2112720 &  24.5183 &  5.0370 &  4.5270 & O 8.0 V&3.1007708 &  24.7019 &  4.9880 &  4.5450 & O 7.5 V\\
 4.9560110 &  24.3579 &  5.0780 &  4.5130 & O 8.5 V&4.1141835 &  24.5674 &  5.0360 &  4.5280 & O 8.0 V\\
 5.1965515 &  24.2999 &  5.0920 &  4.5080 & O 8.5 III&4.7144740 &  24.4782 &  5.0670 &  4.5120 & O 8.5 V\\
 5.6416620 &  24.1845 &  5.1200 &  4.4950 & O 9.0 III&4.9476720 &  24.4422 &  5.0810 &  4.5050 & O 8.5 III\\
 5.9484235 &  24.0992 &  5.1400 &  4.4840 & O 9.5 III&5.1685970 &  24.4077 &  5.0940 &  4.4960 & O 9.0 III\\
 6.3443255 &  23.9827 &  5.1680 &  4.4660 & BSG&5.4422980 &  24.3650 &  5.1110 &  4.4840 & O 9.5 III\\
 8.0793755 &  21.5113 &  5.3640 &  4.1330 & LBV&5.6963615 &  24.3262 &  5.1270 &  4.4690 & BSG\\
 8.0813105 &  21.4602 &  5.3720 &  3.8720 & YHG&6.6092615 &  24.1399 &  5.3040 &  4.0930 & LBV\\
 8.0819755 &  21.4540 &  5.3740 &  3.5900 & RSG&6.6769850 &  23.4420 &  5.3060 &  3.8720 & YHG\\
 8.4990530 &  13.7862 &  5.5290 &  3.8550 & YHG&6.7484280 &  22.9241 &  5.3000 &  3.6760 & RSG\\
 8.5055150 &  13.6556 &  5.5290 &  4.2870 & WNL&7.2963415 &  16.6110 &  5.2570 &  3.5630 & RSG\\
 8.7131440 &  11.4163 &  5.4310 &  4.6900 & WNE&&&&&\\
 8.7183250 &  11.3330 &  5.4520 &  4.7010 & WNE&&&&&\\
\hline
\multicolumn{5}{c|}{$m_\mathrm{ini}$ = 20 $M_\odot$, $v_\mathrm{rot}$
= 300 km/s, $z$ = 0.02}&\multicolumn{5}{c}{$m_\mathrm{ini}$ = 20
  $M_\odot$, $v_\mathrm{rot}$ = 0 km/s, $z$ = 0.02}\\
 0.0000000 &20.0000 &4.6100 &4.5290 &O 8.0 V& 0.0000000 &20.0000 &4.6210 &4.5400 &O 7.5 V\\
 1.2163962 &19.9643 &4.6410 &4.5210 &O 8.5 V& 0.2819594 &19.9930 &4.6260 &4.5360 &O 8.0 V\\
 4.0353365 &19.8546 &4.7370 &4.5060 &O 9.0 V& 2.5817525 &19.9281 &4.6950 &4.5210 &O 8.5 V\\
 5.8747015 &19.7486 &4.8150 &4.4880 &O 9.5 V& 4.6272785 &19.8518 &4.7700 &4.5040 &O 9.0 V\\
 6.7434725 &19.6844 &4.8580 &4.4740 &B 0.0 V& 5.6688175 &19.8048 &4.8140 &4.4890 &O 9.5 V\\
 7.2805585 &19.6392 &4.8870 &4.4630 &BSG& 6.3578040 &19.7711 &4.8470 &4.4740 &B 0.0 V\\
10.1916910 &18.0257 &5.1440 &3.8590 &YG& 6.8873955 &19.7447 &4.8740 &4.4580 &BSG\\
10.1918490 &18.0252 &5.1430 &3.6040 &RSG& 8.4809040 &18.8094 &  5.0720 &  3.8730 &  YG\\
11.0120840 &11.7873 &5.3870 &3.5850 &RSG& 8.5819690 &18.5271 &  5.0450 &  3.6020 & RSG\\
&&&&&9.1767410 &15.7450 &5.0860 &3.5600 &RSG\\
\hline
\hline
\multicolumn{10}{c}{$z$ = 0.04}\\
\hline
\multicolumn{5}{c|}{$m_\mathrm{ini}$ = 120 $M_\odot$,
  $v_\mathrm{rot},$ = 300 km/s, $z$ =
  0.04}&\multicolumn{5}{c}{$m_\mathrm{ini}$ = 120 $M_\odot$,
  $v_\mathrm{rot}$ = 0 km/s, $z$ = 0.04}\\
 0.0000000 & 120.0000 &  6.2810 &  4.7200 & O 2If$^\ast$& 0.0000000 & 120.0000 &  6.2860 &  4.7270 &   O 2If$^\ast$\\
 0.7122990 & 106.1743 &  6.2800 &  4.7040 & WNL&1.1811930 &  96.0022 &  6.2780 &  4.6690 & WNL\\
 1.5903378 &  83.8015 &  6.2800 &  4.3490 & LBV&1.8489928 &  74.8679 &  6.2810 &  4.3330 & LBV\\
 2.0647445 &  41.2997 &  5.9280 &  4.4000 & WNL&2.1893340 &  35.1802 &  5.9270 &  4.4010 & WNL\\
 3.0004895 &  12.6752 &  5.3520 &  4.6570 & WNE&2.5832630 &  16.9112 &  5.5700 &  4.6230 & WNE\\
 3.2045120 &   9.8998 &  5.2310 &  4.6640 &  WC&2.7133730 &  13.5116 &  5.4550 &  4.6370 &  WC\\
 3.4608532 &   7.1121 &  5.1160 &  4.8120 &  WC&3.0063900 &   8.5724 &  5.2650 &  4.8030 &  WC\\
\hline
\multicolumn{5}{c|}{$m_\mathrm{ini}$ = 85 $M_\odot$,
  $v_\mathrm{rot},$ = 300 km/s, $z$ =
  0.04}&\multicolumn{5}{c}{$m_\mathrm{ini}$ = 60 $M_\odot$,
  $v_\mathrm{rot},$ = 300 km/s, $z$ = 0.04}\\
 0.0000000 &  85.0000 &  6.0160 &  4.6700 & O 2.0 V& 0.0000000 &  60.0000 &  5.7630 &  4.6760 & O 3.0 V\\
 0.2073841 &  83.2480 &  6.0170 &  4.6610 & O 2.0 III&1.6115779 &  52.5985 &  5.8280 &  4.6540 & WNL\\
 0.5546225 &  80.1208 &  6.0230 &  4.6450 & O 2.5 III&3.9731632 &  11.5407 &  5.2770 &  4.6680 & WNE\\
 0.6112433 &  79.5865 &  6.0240 &  4.6430 & O 2.5 I&4.2376410 &   8.7119 &  5.1460 &  4.6870 &  WC\\
 0.9455716 &  76.2783 &  6.0320 &  4.6330 & O 3.0 I&4.4505125 &   6.6861 &  5.0990 &  4.8100 &  WC\\
 1.1074332 &  74.5755 &  6.0370 &  4.6310 & WNL&&&&&\\
 2.0882518 &  61.0788 &  6.0530 &  4.3290 & LBV&&&&&\\
 2.5212965 &  38.3726 &  5.8210 &  4.3630 & WNL&&&&&\\
 3.5020968 &  13.2380 &  5.3910 &  4.6520 & WNE&&&&&\\
 3.5656982 &  12.1506 &  5.3470 &  4.6570 &  WC&&&&&\\
 3.9473910 &   7.2946 &  5.1400 &  4.8140 &  WC&&&&&\\
\hline
\multicolumn{5}{c|}{$m_\mathrm{ini}$ = 60 $M_\odot$,
  $v_\mathrm{rot}$ = 0 km/s, $z$ = 0.04}&
\multicolumn{5}{c}{$m_\mathrm{ini}$ = 40 $M_\odot$, $v_\mathrm{rot},$
  = 300 km/s, $z$ = 0.04}\\ 
0.0000000 &  60.0000 &  5.7490 &  4.6630 & O 3.0 V&0.0000000 &  40.0000 &  5.3890 &  4.6190 & O 5.0 V\\
0.2856618 &  59.0542 &  5.7550 &  4.6530 & O 3.5 V&0.2346178 &  39.7772 &  5.3970 &  4.6160 & O 5.5 V\\
0.7580356 &  57.3484 &  5.7690 &  4.6340 & O 4.0 V&1.8258416 &  37.8880 &  5.4750 &  4.5820 & O 6.0 V\\
0.8683556 &  56.9232 &  5.7730 &  4.6290 & O 4.0 III&1.9091198 &  37.7668 &  5.4800 &  4.5800 & O 6.0 III\\
1.1859810 &  55.6416 &  5.7840 &  4.6130 & O 4.5 III&2.6890815 &  36.4932 &  5.5300 &  4.5560 & O 6.5 III\\
1.4319508 &  54.5957 &  5.7940 &  4.5950 & O 5.0 III&3.1912810 &  35.4259 &  5.5660 &  4.5400 & O 7.0 III\\
1.5242232 &  54.1944 &  5.7980 &  4.5880 & O 5.0 I&3.2596885 &  35.2664 &  5.5710 &  4.5370 & WNL\\
1.6160245 &  53.7916 &  5.8020 &  4.5800 & O 5.5 I&4.7375470 &  31.0692 &  5.7210 &  4.2620 & LBV\\
1.8360944 &  52.8212 &  5.8130 &  4.5590 & O 6.0 I&4.7952865 &  29.8853 &  5.8170 &  4.3600 & WNL\\
1.9606289 &  52.2799 &  5.8200 &  4.5430 & O 6.5 I&4.7958090 &  29.8715 &  5.8180 &  4.3480 & LBV\\
2.0813522 &  51.7800 &  5.8270 &  4.5220 & O 7.0 I&4.8024910 &  29.6620 &  5.9030 &  3.8750 & YHG\\
2.2386800 &  51.1887 &  5.8380 &  4.4920 & O 7.5 I&4.8570485 &  27.6540 &  5.9250 &  3.8760 & LBV\\
2.3116478 &  50.9483 &  5.8440 &  4.4750 & O 8.0 I&4.9483710 &  22.8732 &  5.8880 &  4.4330 & WNL\\
2.3481315 &  50.8375 &  5.8470 &  4.4660 & O 8.5 I&5.0142215 &  19.8232 &  5.7240 &  4.5940 & WNE\\
2.3846155 &  50.7335 &  5.8500 &  4.4570 & O 9.0 I&5.0210715 &  19.4902 &  5.7150 &  4.5960 &  WC\\
2.4523712 &  50.5647 &  5.8550 &  4.4330 & O 9.5 I&5.3272680 &  11.4181 &  5.4850 &  4.8340 &  WC\\
2.5535530 &  49.2773 &  5.8580 &  4.3860 & BSG&&&&&\\
2.5865480 &  48.4843 &  5.8570 &  4.3600 & LBV&&&&&\\
3.0187062 &  33.1277 &  5.8510 &  4.3660 & WNL&&&&&\\
3.1266585 &  29.0412 &  5.9810 &  4.4210 & LBV&&&&&\\
3.1425120 &  27.8343 &  6.0130 &  4.4380 & WNL&&&&&\\
3.1872218 &  24.6263 &  5.8430 &  4.5720 & WNE&&&&&\\
3.2445090 &  21.0379 &  5.7580 &  4.5490 &  WC&&&&&\\
3.5380940 &  11.3294 &  5.4760 &  4.8120 &  WC&&&&&\\
\hline
\multicolumn{5}{c|}{$m_\mathrm{ini}$ = 25 $M_\odot$, $v_\mathrm{rot},$
  = 300 km/s, $z$ = 0.04}& \multicolumn{5}{c}{$m_\mathrm{ini}$ = 25
  $M_\odot$, $v_\mathrm{rot},$ = 0 km/s, $z$ = 0.04}\\
 0.0000000 &  25.0000 &  4.9310 &  4.5710 & O 6.5 V&0.0000000 &  25.0000 &  4.9410 &  4.5810 & O 6.5 V\\
 0.7839074 &  24.8724 &  4.9610 &  4.5650 & O 7.0 V&1.8920445 &  24.6451 &  5.0200 &  4.5620 & O 7.0 V\\
 3.4727440 &  24.2181 &  5.1020 &  4.5410 & O 7.5 V&3.2136872 &  24.3080 &  5.0880 &  4.5420 & O 7.5 V\\
 4.7023685 &  23.6354 &  5.1870 &  4.5230 & O 8.0 III&3.9164080 &  24.0910 &  5.1300 &  4.5240 & O 8.0 V\\
 5.5065380 &  23.1187 &  5.2540 &  4.5060 & O 8.5 III&4.1261118 &  24.0218 &  5.1430 &  4.5160 & O 8.0 III\\
 6.1294445 &  22.6320 &  5.3170 &  4.4890 & O 9.0 III&4.2627490 &  23.9758 &  5.1520 &  4.5110 & O 8.5 III\\
 6.4360705 &  22.3614 &  5.3520 &  4.4790 & WNL&4.5867530 &  23.8655 &  5.1750 &  4.4950 & O 9.0 III\\
 7.3809685 &  21.2317 &  5.5980 &  4.2400 & LBV&4.8273150 &  23.7842 &  5.1930 &  4.4800 & O 9.5 III\\
 7.3820465 &  21.0639 &  5.6220 &  3.8730 & YHG&5.0530525 &  23.7106 &  5.2100 &  4.4640 & BSG\\
 7.4232860 &  18.6763 &  5.7080 &  3.8770 & LBV&5.8147580 &  23.5028 &  5.3230 &  4.4940 & O 8.5 III\\
 7.4467310 &  18.1922 &  5.7100 &  4.3290 & WNL&5.8163300 &  23.5015 &  5.3150 &  4.4890 & O 9.0 III\\
 7.5964520 &  14.5465 &  5.5070 &  4.6330 & WNE&5.8163300 &  23.5015 &  5.3150 &  4.4890 & O 9.0 III\\
 7.6086400 &  14.2369 &  5.4950 &  4.6350 &  WC&5.8170285 &  23.5010 &  5.3160 &  4.4750 & O 9.5 III\\
 7.8694710 &   9.5879 &  5.3550 &  4.8140 &  WC&5.8178670 &  23.5004 &  5.3220 &  4.4520 & BSG\\
 &&&&&5.8241865 &  23.4506 &  5.3650 &  4.1410 & LBV\\
 &&&&&5.8282440 &  23.3195 &  5.3910 &  3.8720 & YHG\\
 &&&&&6.4117615 &  13.8709 &  5.4330 &  3.8500 & YHG\\
\hline
\multicolumn{5}{c|}{$m_\mathrm{ini}$ = 20 $M_\odot$, $v_\mathrm{rot},$
  = 300 km/s, $z$ = 0.04}& \multicolumn{5}{c}{}\\
 0.0000000 &  20.0000 &  4.6630 &  4.5210 & O 8.5 V&&&&&\\
 2.3022982 &  19.8656 &  4.7420 &  4.5060 & O 9.0 V&&&&&\\
 3.9294748 &  19.7344 &  4.8150 &  4.4890 & O 9.5 V&&&&&\\
 4.8387095 &  19.6420 &  4.8620 &  4.4740 & B0.0 V&&&&&\\
 5.3234640 &  19.5860 &  4.8890 &  4.4640 & BSG&&&&&\\
 8.1257930 &  17.7537 &  5.0520 &  4.3720 & WNL&&&&&\\
 8.6099200 &  16.8655 &  5.1760 &  3.9950 & BSG&&&&&\\
 8.6106560 &  16.8549 &  5.1750 &  3.8660 &  YG&&&&&\\
 8.7348380 &  15.2882 &  5.3010 &  3.7260 & YHG&&&&&\\
 8.8588410 &  13.6160 &  5.3000 &  3.7630 &  YG&&&&&\\
 9.1548770 &   9.8867 &  5.2440 &  4.4180 & WNL&&&&&\\
 9.4105940 &   9.2449 &  5.3520 &  4.4390 & WNL&&&&&\\
\hline
\hline
\multicolumn{10}{c}{LMC metallicity ($z$ = 0.008)}\\
\hline
\multicolumn{5}{c|}{$m_\mathrm{ini}$ = 120 $M_\odot$,
  $v_\mathrm{rot},$ = 300 km/s, $z$ =
  0.008}&\multicolumn{5}{c}{$m_\mathrm{ini}$ = 60 $M_\odot$,
  $v_\mathrm{rot}$ = 300 km/s, $z$ = 0.008}\\
 0.0000000 & 120.0000 &  6.2140 &  4.7310 & O 2.0 If$^\ast$&0.0000000 &  60.0000 &  5.6820 &  4.6860 & O 3.0 V\\
 1.9595181 & 102.9785 &  6.3320 &  4.6760 & WNL&0.2979827 &  59.6762 &  5.6950 &  4.6770 & O 3.5 V\\
 2.5262850 &  93.5009 &  6.3860 &  4.3640 & LBV&0.7291670 &  59.1504 &  5.7170 &  4.6700 & O 3.5 III\\
 3.1879975 &  37.2673 &  6.0530 &  4.4570 & WNL&2.0456922 &  56.9579 &  5.7980 &  4.6420 & O 4.0 III\\
 3.2485625 &  33.3330 &  6.0390 &  4.5960 & WNE&2.3903512 &  56.1718 &  5.8240 &  4.6310 & O 4.0 I\\
 3.2724660 &  31.8841 &  6.0180 &  4.5790 &  WC&2.6550700 &  55.4854 &  5.8450 &  4.6210 & O 4.5 I\\
 3.2890230 &  30.4860 &  5.9900 &  4.4240 & LBV&3.0891092 &  54.1722 &  5.8840 &  4.6000 & O 5.0 I\\
 3.2919092 &  30.0792 &  5.9740 &  4.4960 &  WC&3.4820080 &  52.7520 &  5.9230 &  4.5800 & O 5.5 I\\
 3.6300872 &  13.3927 &  5.5950 &  4.8180 &  WC&3.8287820 &  51.2107 &  5.9620 &  4.5620 & WNL\\
&&&&& 4.4030205 &  48.4254 &  6.0750 &  4.4590 & LBV\\
&&&&& 4.4045680 &  47.8923 &  6.1060 &  4.5100 & WNL\\
&&&&& 4.4048185 &  47.8903 &  6.1080 &  4.4800 & LBV\\
&&&&& 4.4081185 &  46.8239 &  6.1460 &  3.8770 & BSG\\
&&&&& 4.4090990 &  46.3901 &  6.1530 &  3.8750 & YHG\\
&&&&& 4.4110175 &  45.4902 &  6.1600 &  3.8770 & LBV\\
&&&&& 4.5355055 &  36.0793 &  6.1300 &  4.5000 & WNL\\
&&&&& 4.6174305 &  31.6506 &  6.0390 &  4.6050 & WNE\\
&&&&& 4.6207880 &  31.4079 &  6.0340 &  4.5980 &  WC\\
&&&&& 4.6518545 &  29.8221 &  6.0070 &  4.2280 & LBV\\
&&&&& 4.6602595 &  26.1187 &  5.9310 &  4.4170 &  WC\\
&&&&& 4.8432090 &  16.4456 &  5.7330 &  4.8490 &  WC\\
\hline
\multicolumn{5}{c|}{$m_\mathrm{ini}$ = 40 $M_\odot$,
  $v_\mathrm{rot},$ = 300 km/s, $z$ =
  0.008}&\multicolumn{5}{c}{$m_\mathrm{ini}$ = 30 $M_\odot$,
  $v_\mathrm{rot}$ = 300 km/s, $z$ = 0.008}\\
 0.0000000 &  40.0000 &  5.3170 &  4.6430 & O 5.5 V&0.0000000 &  30.0000 &  5.0330 &  4.6080 & O 6.5 V\\
 2.3276098 &  39.0577 &  5.4390 &  4.6150 & O 5.5 III&0.2138163 &  29.9785 &  5.0390 &  4.6050 & O 7.0 V\\
 3.1412225 &  38.5397 &  5.4930 &  4.5980 & O 6.0 III&3.2977295 &  29.5309 &  5.1860 &  4.5780 & O 7.5 V\\
 3.7145050 &  38.0821 &  5.5360 &  4.5780 & O 6.5 III&3.9570908 &  29.3825 &  5.2260 &  4.5670 & O 7.5 III\\
 4.0200750 &  37.8021 &  5.5610 &  4.5640 & O 6.5 I&5.0176330 &  29.0838 &  5.2990 &  4.5380 & O 8.0 III\\
 4.0936185 &  37.7309 &  5.5680 &  4.5600 & O 7.0 I&5.4466060 &  28.9076 &  5.3320 &  4.5190 & O 8.5 III\\
 4.4425960 &  37.3758 &  5.5990 &  4.5390 & O 7.5 I&5.7691770 &  28.7612 &  5.3590 &  4.5010 & O 9.0 I\\
 4.6363970 &  37.1147 &  5.6180 &  4.5240 & O 8.0 I&6.0574980 &  28.6264 &  5.3850 &  4.4800 & O 9.5 I\\
 4.8845875 &  36.7766 &  5.6430 &  4.5030 & O 8.5 I&6.2608615 &  28.5321 &  5.4040 &  4.4620 & BSG\\
 5.0001165 &  36.6231 &  5.6550 &  4.4890 & O 9.0 I&6.9819635 &  27.4565 &  5.5050 &  4.4630 & O 9.5 I\\
 5.1102935 &  36.4814 &  5.6670 &  4.4730 & O 9.5 I&6.9829310 &  27.4558 &  5.5020 &  4.4610 & BSG\\
 5.2633395 &  36.2980 &  5.6850 &  4.4480 & BSG&6.9879585 &  27.4354 &  5.5310 &  4.2160 & LBV\\
 5.6510235 &  35.2933 &  5.7580 &  4.4830 & O 9.0 I&6.9905500 &  27.3408 &  5.5430 &  3.8650 & YHG\\
 5.6524505 &  35.2902 &  5.7560 &  4.4680 & O 9.5 I&7.3154990 &  16.2920 &  5.6210 &  4.4620 & WNL\\
 5.6528645 &  35.2895 &  5.7580 &  4.4500 & BSG&7.5867505 &  13.6902 &  5.5440 &  4.6940 & WNE\\
 5.6533780 &  35.2868 &  5.7650 &  4.4110 & BSG&7.5963590 &  13.5216 &  5.5390 &  4.6890 &  WC\\
 5.6541770 &  35.2744 &  5.7780 &  4.3290 & LBV&7.6507400 &  12.1059 &  5.5420 &  4.6590 &  WC\\
 5.6566830 &  35.0328 &  5.8000 &  3.8750 & YHG&&&&&\\
 5.7411655 &  28.4731 &  5.8650 &  3.8780 & LBV&&&&&\\
 5.7845300 &  27.3243 &  5.8730 &  4.3750 & WNL&&&&&\\
 5.8047500 &  26.9017 &  5.8720 &  4.3660 & LBV&&&&&\\
 5.8336760 &  26.2948 &  5.8700 &  4.3730 & WNL&&&&&\\
 6.1551555 &  19.5713 &  5.7860 &  4.6640 &  WC&&&&&\\
 6.2103525 &  17.3421 &  5.7650 &  4.7800 &  WC&&&&&\\
\hline
\multicolumn{5}{c|}{$m_\mathrm{ini}$ = 20 $M_\odot$, $v_\mathrm{rot},$
  = 300 km/s, $z$ = 0.008, Padova model}&\multicolumn{5}{c}{}\\
 0.0000000 &  19.6947 &  4.6294 &  4.5613 & O  8.5 V&&&&&\\
 1.5522020 &  19.6766 &  4.6652 &  4.5470 & O  9.0 V&&&&&\\
 5.3253110 &  19.6087 &  4.8051 &  4.5263 & O  9.5 V&&&&&\\
 7.2242600 &  19.5097 &  4.8985 &  4.4929 & B&&&&&\\
 7.5282720 &  19.4872 &  4.9155 &  4.4837 & BG&&&&&\\
 9.0765410 &  19.2730 &  5.0680 &  4.3931 & BSG&&&&&\\
 9.0885540 &  19.2641 &  5.0865 &  3.8735 &  YG&&&&&\\
 9.0919190 &  19.2641 &  5.0451 &  3.6387 & RSG&&&&&\\
 9.1301450 &  19.2243 &  5.1754 &  3.6576 &  YG&&&&&\\
 9.1327400 &  19.2243 &  5.1716 &  3.8853 & BSG&&&&&\\
 9.5183220 &  19.0173 &  5.1648 &  3.8710 &  YG&&&&&\\
 9.5882510 &  18.9692 &  5.1509 &  3.6502 & RSG&&&&&\\
 9.7910080 &  18.7219 &  5.2919 &  3.5601 & RSG&&&&&\\
\hline
\hline
\multicolumn{10}{c}{SMC metallicity ($z$ = 0.004)}\\
\hline
\multicolumn{5}{c|}{$m_\mathrm{ini}$ = 120 $M_\odot$,
  $v_\mathrm{rot},$ = 300 km/s, $z$ =
  0.004}&\multicolumn{5}{c}{$m_\mathrm{ini}$ = 60 $M_\odot$,
  $v_\mathrm{rot},$ = 300 km/s, $z$ =
  0.004}\\
 0.0000000 & 120.0000 &  6.2090 &  4.7450 & O 2.0 If$^\ast$&0.0000000 &  60.0000 &  5.6760 &  4.6980 & O 2.0 If$^\ast$\\
 2.2878100 & 104.9922 &  6.3760 &  4.6680 & WNL&1.6283401 &  58.4894 &  5.7700 &  4.6700 & O 2.0 V\\
 2.7973842 &  95.8416 &  6.4420 &  4.3910 & LBV&2.2080835 &  57.6779 &  5.8110 &  4.6540 & O 2.5 III\\
 3.2331478 &  52.7831 &  6.3080 &  4.5710 & WNL&2.6747342 &  56.8529 &  5.8490 &  4.6370 & O 3.0 III\\
 3.2714685 &  49.1669 &  6.2730 &  4.5860 & WNE&2.9854308 &  56.1942 &  5.8760 &  4.6200 & O 3.5 III\\
 3.2810722 &  48.1992 &  6.2630 &  4.5220 & LBV&3.1044190 &  55.9151 &  5.8880 &  4.6130 & O 4.0 I\\
 3.3344045 &  34.8949 &  6.0790 &  4.4690 &  WC&3.2770062 &  55.4829 &  5.9050 &  4.6010 & O 4.5 I\\
 3.5980968 &  17.1781 &  5.7530 &  4.7950 &  WC&3.4442515 &  55.0331 &  5.9220 &  4.5880 & O 5.0 I\\
\cline{1-5}\multicolumn{5}{c|}{$m_\mathrm{ini}$ = 40 $M_\odot$,
  $v_\mathrm{rot},$ = 300 km/s, $z$ = 0.004}&3.5500428 &  54.7333 &  5.9340 &  4.5790 & O 5.5 I\\
 0.0000000 &  40.0000 &  5.3120 &  4.6550 & O 3.5 V&3.7019615 &  54.2844 &  5.9510 &  4.5650 & O 6.0 I\\
 0.6528106 &  39.8576 &  5.3410 &  4.6460 & O 4.0 V&3.8458675 &  53.8327 &  5.9680 &  4.5480 & O 6.5 I\\
 2.5598660 &  39.2560 &  5.4490 &  4.6270 & O 4.5 V&3.9812960 &  53.3258 &  5.9840 &  4.5340 & O 7.0 I\\
 3.3546038 &  38.8721 &  5.5050 &  4.6090 & O 5.0 V&4.1079905 &  52.8620 &  6.0000 &  4.5170 & O 7.5 I\\
 3.6820315 &  38.6794 &  5.5300 &  4.5990 & O 5.0 III&4.1855345 &  52.5885 &  6.0110 &  4.5030 & O 8.0 I\\
 3.8373135 &  38.5797 &  5.5420 &  4.5930 & O 5.5 III&4.2578970 &  52.3466 &  6.0200 &  4.4870 & O 8.5 I\\
 4.2053280 &  38.3209 &  5.5730 &  4.5760 & O 6.0 III&4.3262125 &  52.1321 &  6.0300 &  4.4720 & O 9.0 I\\
 4.4137455 &  38.1603 &  5.5920 &  4.5630 & O 6.5 III&4.3305155 &  52.1190 &  6.0310 &  4.4710 & WNL\\
 4.5455730 &  38.0540 &  5.6050 &  4.5540 & O 6.5 I&4.4243485 &  51.8059 &  6.0720 &  4.4590 & LBV\\
 4.6090715 &  38.0016 &  5.6110 &  4.5490 & O 7.0 I&4.4274385 &  51.2078 &  6.1040 &  3.8750 & YHG\\
 4.7954600 &  37.8387 &  5.6290 &  4.5320 & O 7.5 I&4.4443775 &  44.5912 &  6.1620 &  4.2180 & LBV\\
 4.9154035 &  37.7017 &  5.6420 &  4.5210 & O 8.0 I&4.6386520 &  35.9884 &  6.1270 &  4.5490 & WNL\\
 5.0891035 &  37.5044 &  5.6600 &  4.5020 & O 8.5 I&4.6752050 &  34.2691 &  6.0890 &  4.2940 & LBV\\
 5.1982725 &  37.3843 &  5.6720 &  4.4870 & O 9.0 I&4.6875580 &  34.2642 &  6.0950 &  4.5950 & WNL\\
 5.3033105 &  37.2729 &  5.6840 &  4.4700 & O 9.5 I&4.7098275 &  33.9055 &  6.0960 &  4.4550 & LBV\\
 5.3987365 &  37.1772 &  5.6950 &  4.4530 & BSG&4.7404865 &  32.5838 &  6.0880 &  4.6610 & WNL\\
 5.6899025 &  36.7542 &  5.7570 &  4.4720 & O 9.0 I&4.7501970 &  32.0460 &  6.0800 &  4.6420 & WNE\\
 5.6909885 &  36.7519 &  5.7550 &  4.4670 & O 9.5 I&4.7551630 &  31.7438 &  6.0760 &  4.6310 & WC\\
 5.6918035 &  36.7506 &  5.7550 &  4.4480 & BSG&4.8120270 &  28.4654 &  6.0630 &  4.8410 & WC\\
 5.6937425 &  36.7311 &  5.7740 &  4.3280 & LBV&&&&&\\
 5.6965865 &  36.5473 &  5.7940 &  3.8750 & YHG&&&&&\\
 5.8482340 &  28.0283 &  5.8640 &  3.8760 & LBV&&&&&\\
 5.9362130 &  26.1314 &  5.8730 &  4.4130 & WNL&&&&&\\
 6.1748640 &  22.3331 &  5.8980 &  4.3890 & WNL&&&&&\\
\hline
\multicolumn{5}{c|}{$m_\mathrm{ini}$ = 20 $M_\odot$,
  $v_\mathrm{rot},$ = 300 km/s, $z$ =
  0.004, Padova model}&\multicolumn{5}{c}{$m_\mathrm{ini}$ = 15 $M_\odot$,
  $v_\mathrm{rot},$ = 300 km/s, $z$ =
  0.004, Padova model}\\
 0.0000000 &  19.6947 &  4.6301 &  4.5733 & O 7.5 V&0.0000000 &  14.7724 &  4.2919 &  4.5249 & O 9.5 V\\
 0.6778070 &  19.6947 &  4.6479 &  4.5648 & O 8.0 V&6.9185840 &  14.7622 &  4.4615 &  4.5012 & B\\
 5.5389360 &  19.6268 &  4.8283 &  4.5412 & O 8.5 V&11.9532000 &  14.7214 &  4.6655 &  4.4266 & BSG\\
 6.5944990 &  19.5997 &  4.8804 &  4.5255 & O 9.0 V&12.7442700 &  14.7045 &  4.7348 &  4.4348 & B\\
 7.3414660 &  19.5591 &  4.9215 &  4.5078 & O 9.5 V&12.7579800 &  14.7045 &  4.7453 &  4.4276 & BSG\\
 7.8724170 &  19.5276 &  4.9537 &  4.4893 & B&12.7770200 &  14.7045 &  4.7461 &  3.8726 & YG\\
 8.2688160 &  19.5007 &  4.9807 &  4.4698 & BG&12.7791200 &  14.7045 &  4.6056 &  3.6517 & RSG\\
 8.9886990 &  19.4111 &  5.0836 &  4.4244 & BSG&12.9364100 &  14.6639 &  4.8698 &  3.6534 & YG\\
 9.6339610 &  19.1447 &  5.1912 &  3.8731 & YG&12.9389100 &  14.6639 &  4.8960 &  3.8775 & BSG\\
 9.6607520 &  19.1227 &  5.1664 &  3.6523 & RSG&13.7597900 &  14.5764 &  4.7932 &  3.8694 & YG\\
&&&&&13.7622400 &  14.5764 &  4.6591 &  3.6501 & RSG\\
\hline
\multicolumn{10}{c}{\tablefoot{
Evolution of the \citet{MM03}
  and \citet{MM05} stellar models with and without rotation through
  the spectral types with a metallicity of $z$ = 0.02, 0.04, 0.008 and
  0.004. The 20 $M_\odot$ model with z = 0.008 and the
    15 and 20 $M_\odot$ models with z = 0.004 are from \citet{BNG09}.}}
\end{longtable}

\label{lastpage}
\end{document}